\documentclass[11pt]{article}
\usepackage{times}
\usepackage{colordvi}
\usepackage{graphics}
\usepackage{makeidx}
\usepackage{epsf}
\usepackage{here}
\usepackage{cite}
\usepackage{amsmath}
\usepackage{pstricks}
\usepackage{pst-node}

\newcommand{\nll}{\nonumber \\}

\newcommand{\bq}{\begin{equation}}
\newcommand{\eq}{\end{equation}}
\newcommand{\ba}{\begin{eqnarray}}
\newcommand{\ea}{\end{eqnarray}}
\def \3{\ss} 
\newcommand{\nobody}{\rule{0ex}{1ex}}

\newcommand{\nobodyfrac}{\frac{\nobody}{\nobody}}

\numberwithin{equation}{section}
\numberwithin{figure}{section}
\numberwithin{table}{section}

\begin{document}

\def\sigmap{\sigma^{\prime}}
\def\mup{\mu^{\prime}}
\def\nup{\nu^{\prime}}
\def\rhop{\rho^{\, \prime}}
\def\bb{b \bar{b}} 
\def\cc{c \bar{c}} 
\def\qq{q \bar{q}} 
\def\cM{{\cal M}} 
\def\cO{{\cal O}}
\def\cK{{\cal K}} 
\def \ni {\noindent}
\def \be {\begin{equation}}
\def \e {\end{equation}}
\def \bea {\begin{eqnarray}}
\def \ea {\end{eqnarray}}
\def \eps {\epsilon}
\def \si {\sigma}
\def \ga {\gamma}
\def \ka {\kappa}
\def \la {\lambda}
\def \no {\nonumber}
\def \G {{\rm g}}
\def \dd {{\rm d}}
\def \Li {{\rm Li_2}}
\def \K {k_{\bot}^2}
\def \Vec#1{\mbox {\bf #1}}
\def \Veg#1{\mbox{\boldmath $#1$}}
\def \sub {\scriptscriptstyle}
\def \ps {p\hspace{-0.43em}/}
\def \mps {p\hspace{-0.45em}/}
\def \sps {p\hspace{-0.32em}/}
\def \ns {n\hspace{-0.51em}/}
\def \mns {n\hspace{-0.53em}/}
\def \ks {k\hspace{-0.49em}/}
\def \es {\epsilon\hspace{-0.47em}/}
\def \Z#1#2{\widetilde{Z}^{#1}_{#2}}
\def \Zw{\ensuremath{ {Z^{-\frac{1}{2}}_{\scriptscriptstyle W}\, } }}
\def \Zz{\ensuremath{ {Z^{-\frac{1}{2}}_{\scriptscriptstyle Z}\, } }}
\def \Zwn{\ensuremath{ \delta Z_{\scriptscriptstyle W,\,n} \,}}
\def \Zzn{\ensuremath{ \delta Z_{\scriptscriptstyle Z,\,n} \,}}
\def \slash#1{#1 \hspace{-0.42em}/}
\newcommand{\To}[2]{\stackrel{#1}{\hbox to #2 pt{\rightarrowfill}}}

\def \an {\widehat}
\def \abs#1{|\,#1\,|}
\def \vector#1{\stackrel{\hspace{-0.45em}\longrightarrow}{#1}}
\def \pa {\partial}
\def \c {\hspace{-0.2em} \cdot}
\def\wp{\ifmmode W^+\else $W^+$\fi}
\def\wm{\ifmmode W^-\else $W^-$\fi}
\def\emm{\ifmmode e^-\else $e^-$\fi}
\def\ep{\ifmmode e^+\else $e^+$\fi}

\def\sw{\ensuremath{ \sin \theta_{\rm w}}}
\def\swto{\ensuremath{ \sin^2 \theta_{\rm w}}} 
\def\swfor{\ensuremath{ \sin^4 \theta_{\rm w}}} 
\def\cw{\ensuremath{ \cos \theta_{\rm w}}} 
\def\cwto{\ensuremath{ \cos^2 \theta_{\rm w}}}
\def\cwfor{\ensuremath{ \cos^4 \theta_{\rm w}}}

\def\ie{{\it i.e.~}}
\def\eg{{\it e.g.~}}

\def\tw{\ensuremath{ \tan \theta_{\rm w} }}
\def\ctw{\ensuremath{ \cot \theta_{\rm w} }}
\def\twto{\ensuremath{ \tan^2 \theta_{\rm w} }}
\def\ctwto{\ensuremath{ \cot^2 \theta_{\rm w} }}

\def\mw{\ensuremath{ {M}_{\scriptscriptstyle W}} }
\def\mz{\ensuremath{ {M}_{\scriptscriptstyle Z}} }
\def\mh{\ensuremath{ {M}_{\scriptscriptstyle H}} }
\def\mn{\ensuremath{ {M}_{\scriptscriptstyle N}} }
\def\mwp{\ensuremath{ {M}_{{\scriptscriptstyle W},\,{\rm phys. }}}}
\def\mzp{\ensuremath{ {M}_{{\scriptscriptstyle Z},\,{\rm phys. }}}}

\def \mt{\ensuremath{ m_t}}

\def\B#1{\ensuremath{\delta Z^{\, (1)}_{W_T}({#1})   }}
\def\S#1{\ensuremath{\delta Z^{\, (1)}_{\phi}({#1})   }}

\def \Cw#1{\ensuremath { \delta Z^{\, (1)}_{f_{#1}}(W)}}
\def \Cz#1{\ensuremath { \delta Z^{\, (1)}_{f_{#1}}(Z)}}
\def \Cga#1{\ensuremath { \delta Z^{\, (1)}_{f_{#1}}(\ga) }}

\def\L{\ensuremath{ {\rm L}}}
\def\z#1{\ensuremath { \frac{\dd z_{#1}}{z_{#1}}  }}
\def\y#1{\ensuremath { \frac{\dd y_{#1}}{y_{#1}}  }}

\def \onshellm{\vphantom{\frac{1}{1}}_{\left|_{\scr{\,p^2=m^2}}\right.}}
\def \onshell{\vphantom{\frac{1}{1}}_{\left|_{\scr{\,k^2=\mw^2}}\right.}}
\def \onshellW{\vphantom{\frac{A^2}{A^2}}\left|_{_{\scr{\,k^2=\mw^2}}}\right.}
\def \onshellZ{\vphantom{\frac{A^2}{A^2}}\left|_{_{\scr{\,p^2=\mz^2}}}\right.}
\def \onshellA{\vphantom{\frac{A^2}{A^2}}\left|_{_{\scr{\,p^2=0}}}\right.}
\def \onshellAIR{\vphantom{\frac{A^2}{A^2}}\left|_{_{\scr{\,k^2 \to 0}}}\right.}

\def \onshellH{\vphantom{\frac{A^2}{A^2}}\left|_{_{\scr{\,k^2=\mh^2}}}\right.}
\def \onshellf{\vphantom{\frac{A^2}{A^2}}\left|_{_{\scr{\,p^2=m_f^2}}}\right.}
\def \onshellp{\vphantom{\frac{1}{1}}_{\left|_{\scr{\,k^2=\mwp^2}}\right.}}
\def \onshellWp{\vphantom{\frac{A^2}{A^2}}\left|_{_{\scr{\,k^2=\mwp^2}}}\right.}

\def \dd#1{\frac{\partial}{\partial \, {#1}^2}}

\def \h#1{ \hspace*{#1mm}}
\def \v#1{ \vspace*{#1mm}}

\def\dis#1{\ensuremath { {\displaystyle  #1}}}  
\def\scr#1{\ensuremath { { \scriptstyle #1}}}  
\def\sscr#1{\ensuremath { { \scriptscriptstyle #1}}}  

\def \myto#1#2{\ensuremath {\begin{array}[H]{c}
 {\scriptstyle {\rm #1}} \\[-2mm] {-\!\!\!-\!\!\!-\!\!\!-\!\!\!\longrightarrow} \\[-2mm] {\scriptstyle {\rm #2}}
\end{array} }}

\def \myk{\ensuremath {\begin{array}[H]{c} \\[-7mm] {-\!\!\!-\!\!\!\longrightarrow} \\[-2mm] k
\end{array} }}

\def \Myto#1{\ensuremath {\begin{array}[H]{c}
 {\scriptstyle { #1}} \\[-2mm] {-\!\!\!\longrightarrow} \end{array} }}
\def \Myeq#1{\ensuremath {\begin{array}[H]{c}
 {\scriptstyle { #1}} \\[-1mm] {=\!\!\!=\!\!\!=\!\!\!} \end{array} }}

\def \pso {p\hspace{-0.43em}/ _1}
\def \pst {p\hspace{-0.43em}/ _2}
\def \ksp {k\hspace{-0.49em}/_+}
\def \ksm {k\hspace{-0.49em}/_-}
\def \kso {k\hspace{-0.49em}/_1}
\def \kst {k\hspace{-0.49em}/_2}

\def \uo{\ensuremath { u(p_1) \, }}
\def \ut{\ensuremath { \bar{u}(-p_2) \, }}
\def \vo{\ensuremath { v(-p_3) \, }}
\def \vt{\ensuremath { \bar{v}(p_4) \, }}

\def\m{\ensuremath} 

\def \co#1{\ensuremath{ [ (k - q_{#1})^2 ]}}
\def \c#1{\ensuremath{ [ (k - q_{#1})^2 - m_{#1}^2 ]}}
\def \ck#1{\ensuremath{ [ k^2 - m_{#1}^2 ]}}

\def \FP#1#2{\ensuremath { \frac{ i}{[\,{#1} ({#2})-m+i\,\eps\,]} \,\,}}
\def \PP#1#2#3{\ensuremath { \frac{(- i \h{0.5})\,  \G_{#1 #2}}{[\,{#3}^2+i\eps\,]} \,\,}}
\def \BP#1#2#3#4{\ensuremath {\,\frac{ (-i\,e) \,  \G_{#1 #2}}{[\,{#3}^2-{#4}^2+i\,\eps\,]} \,\, }}

\def \Ve#1{\ensuremath { (-i\,e \, Q_e \, \ga^{#1} ) \,\, }}
\def \Vt#1{\ensuremath { (-i\,e \, Q_t \,  \ga^{#1} ) \,\, }}

\def \Vee#1{\ensuremath { (i\,e \, v_e \, \ga^{#1} ) \,\, }}
\def \Vtt#1{\ensuremath { (i\,e \, v_t \,  \ga^{#1} ) \,\, }}

\def \Vze#1{\ensuremath { \big(i\,e \, (v_e + a_e\, \ga_5)\,
\ga^{#1} \big) \,\, }}
\def \Vzt#1{\ensuremath { \big(i\,e \, (v_t + a_t\, \ga_5)\, \ga^{#1} \big) \,\, }}
\def \Vzeo#1{\ensuremath { \big(i\,e \, (v_e^0 + a_e^0\, \ga_5)\,
\ga^{#1} \big) \,\, }}
\def \Vzto#1{\ensuremath { \big(i\,e \, (v_t^0 + a_t^0\, \ga_5)\,
\ga^{#1} \big) \,\, }}

\def \Vzere#1{\ensuremath { \big(i\,e \, (\tilde{v_e})\,
\ga^{#1} \big) \,\, }}
\def \Vztre#1{\ensuremath { \big(i\,e \, (\tilde{v_t})\,
\ga^{#1} \big) \,\, }}

\def \Vw#1{\ensuremath { \left( \frac{i\,e }{ \sqrt{2} \, \sw} \, \ga^{#1} \right) \,\, }}

\def \BD#1#2{\ensuremath {\frac{(-i)}{[\,{#1}^2-{#2}^2+i\,\eps\,]} \,\, }}


\def \Vpww#1#2#3#4#5#6#7{\ensuremath { ({#1} i\,e)\, \Big[
    (-{#6}+{#7})^{\,#2} \G^{\,{#3}\,{#4}} -   ({#7}+{#5})^{\,#3} \,
    \G^{\,{#2}\, {#4}} + ({#5}+{#6})^{\,#4} \, \G^{\,{#2} \, {#3}} \Big]\,\,  }}

\def \Vzww#1#2#3#4#5#6#7{\ensuremath { \left(\frac{{#1} -  i\,e \, \cw}{\sw}\right)\, \Big[
    (-{#6}+{#7})^{\,#2} \G^{\,{#3}\,{#4}} -   ({#7}+{#5})^{\,#3} \,
    \G^{\,{#2}\, {#4}} +  ({#5}+{#6})^{\,#4} \, \G^{\,{#2} \, {#3}} \Big]\,\,  }}

\def \e#1#2{\ensuremath { \eps_{\,#1}^{\,*}\,({#2})\,\, }}

\def \BPCu#1#2#3#4{\ensuremath {\frac{(-i)}{[\,{#3}^2-{#4}^2+i\,\eps\,]}\,
    \left( \, \G^{\, {#1}\, {#2}} + \frac{{#3}_{#1}\,
        {#3}_{#2}}{\vec{#3}^2} - \frac{{#3}_0\, [{#3}_{#1}\,
        n_{#2}+{#3}_{#2}\,n_{#1}\, ]}{\vec{#3}^2} \right)  \,\, }}

\def \eett {\ensuremath {e^+e^- \to t \bar{t} \,\,}}
\def \GeV {\ensuremath  \,{\rm GeV} \,\,}

\def\sig{\left[\frac{\displaystyle{\mathrm{d}\sigma}}{\displaystyle{\mathrm{d}\cos \, \theta}}\right]}

\def\unity{{\rm 1\mskip-4.25mu l}}
\def\re{\mathop{\mathrm{Re}}}

\newcommand{\eqir}{\stackrel{{ \rm IR }}{\longrightarrow}}


\renewcommand{\thefootnote}{\fnsymbol{footnote}}  
\ni
DESY 02--204           \hfill  ISSN 0418--9833
\\  
BI--TP--2002/29
\\
CERN-TH/2003-023
\\
LMU--12/02
\\
SFB/CPP-03-02
\\
hep-ph/0302259
\\
\vspace{1cm} 
\begin{center}  
{\LARGE \bf Elektroweak one-loop corrections for \boldmath{$e^+e^-$} annihilation into
  \boldmath{ $t\bar{t}$} \\[3mm]

 including hard bremsstrahlung}
\footnote{
Work supported in part by the European Community's Human Potential
Programme under contract HPRN-CT-2000-00149 `Physics at Colliders'
 and by Sonderforschungsbereich/Transregio 9 of DFG
`Computergest{\"u}tzte Theoretische Teilchenphysik'. 
This
research has also been supported by a Marie Curie Fellowship of the
European Community Programme `Improving Human Research Potential and
the Socio-economic Knowledge Base' under the contract number HPMF-CT-2002-01694.
}
\\ 
\vspace{1.5cm} 
{ 
{\Large J. Fleischer}${}^{1}$
}, 
~~{\Large A. Leike}${}^{2}$,
~~{\Large T. Riemann}${}^{3}$
~~ and
~~{\Large A. Werthenbach}${}^{3, 4}$%
\footnote{E-mails:~%
fleischer@physik.uni-bielefeld.de,
leike@theorie.physik.uni-muenchen.de,
Tord.Riemann@desy.de,\\
\phantom{E-mails:~~~~~~~~}Anja.Werthenbach@cern.ch
} \\
\vspace{1cm} 

{
{${}^{1}$~Fakult\"at f\"ur Physik, Universit\"at Bielefeld, Universit\"atsstr. 25,  33615
Bielefeld, Germany\\ }
\smallskip
{${}^{2}$~%
Sektion Physik der Universit\"at M\"unchen, Theresienstr. 37, 80333 Munich, Germany
\\ }
\smallskip
{${}^{3}$~Deutsches Elektronen-Synchrotron, DESY Zeuthen, Platanenallee
  6, 15738 Zeuthen, Germany \\ }  
\smallskip
{${}^{4}$~Theory Division, CERN, 1211 Geneva 23, Switzerland }  
}
\end{center}

\vspace{1cm} 

\begin{center}
{\Large \bf 
{Abstract}}
\end{center}
We present the complete electroweak one-loop corrections to top-pair
production at a linear $e^+e^-$ col\-li\-der in the continuum region.
Besides weak and photonic virtual corrections, 
real hard bremsstrahlung with simple realistic kinematical cuts is
included.  For the bremsstrahlung
we  advocate a semi-analyti\-cal approach with a high numerical accuracy.
The virtual corrections are parametrized through six independent form
factors, suitable for Monte-Carlo implementation.
Alternatively, our numerical package {\tt topfit}, 
a stand-alone code, can be utilized
for the calculation of both differential and integrated cross sections as well as
forward--backward asymmetries.

\setcounter{footnote}{0}  
\renewcommand{\thefootnote}{\arabic{footnote}}  
\thispagestyle{empty} 


           \newpage

           \clearpage
\section{Introduction}
\label{chap:intro}
At a future linear $e^+e^-$ collider with a centre-of-mass energy
above 350 GeV, one of the most important {reactions will be
top-pair production well above the threshold} (i.e. in the
continuum region),
\begin{eqnarray}
\label{eq1}
  \label{eq:qqtt}
  e^+ ~+~ e^- \rightarrow t ~+~  \bar t~ \,\,.
\end{eqnarray}
Several hundred thousand events are expected, and the anticipated
accuracy of the corresponding theoretical
predictions should be around a few per mille. 
Of course, it is not only
the two-fermion production process (\ref{eq:qqtt}), with electroweak
radiative corrections (EWRC) and
QCD corrections to the final state that {has} to be calculated with
high precision. Additionally the decay
of the top 
quarks and a variety of quite different radiative corrections such as
real photonic bremsstrahlung and other non-factorizing contributions
to six-fermion production and beamstrahlung  have to be considered.  
Potentially, new physics effects also have to be taken into
account.
For more details on the general subject 
{of top physics, we refer the reader to \cite{Simmons:2002zi} and,} 
for top-pair production, to the recent collider studies 
\cite{Accomando:1997wt,%
Aguilar-Saavedra:2001rg,%
Abe:2001wn} and 
 references therein.

The electroweak one-loop corrections will be a central building
block in any precision study of top-pair production.
 Also it  might well be that for most of the  physics 
 a phenomenological study
of the two-particle (top-pair) production cross section will be
sufficient, thus avoiding to deal too much with many-particle final
states observed in the detectors
\cite{Biernacik:2001mp,Kolodziej:2001xe,Biernacik:2002tt,%
Nekrasov:2002mw,Dittmaier:2002ap}.
For these reasons,
we recalculated the complete set of electroweak contributions,
including real hard photon corrections.
{Several studies on this topic are already available in the literature.
In
\cite{Fujimoto:1988hu,Yuasa:1999rg}, 
the complete $O(\alpha)$ corrections, including hard photon radiation,
 are calculated.
The virtual and soft photon corrections both in the Standard Model and
in the minimal supersymmetric Standard Model are determined in 
\cite{Beenakker:1991ca,Hollik:1998md}, and (only) in the    Standard Model
in
\cite{Bardin:2000kn}.
Experience {proves} that  so far it was difficult
 to get a satisfactory
numerical comparison based on articles or
computer codes without contacting the corresponding authors.
 Due to the importance of the process, for future applications,
it is therefore necessary to provide a common basis and accessible
documentation. 
Thus we aim, with 
the present write-up, to carefully document the one-loop
radiative corrections  for the process 
(\ref{eq1})~\footnote{
The situation with massless fermion-pair production is much better
due to the efforts related to LEP physics; see
\cite{Bardin:1997xq,Kobel:2000aw} and the references therein. 
}}, with the publicly  available 
Fortran program {\tt topfit} \cite{FRW:2002sw,Fleischer:2002kg}, and  with the sample
Fortran outputs.
{In the mean time, we  compared our calculations in detail with the  
results of two collaborations 
\cite{Fleischer:2002rn,Fleischer:2002nn}.\footnote{Another series of
  numerical comparisons with the authors of \cite{Bardin:2000kn} 
was started in September 2001; see also \cite{Andonov:2002xc}.} }

{In this article, 
we sketch in short our calculation and present some typical
numerical results applicable at typical Linear Collider energies. }


\section{Conventions and Cross Sections}
\label{chap:born}
\begin{figure}[tbh]
\begin{center}
\mbox{\epsfysize=3.0cm\epsffile{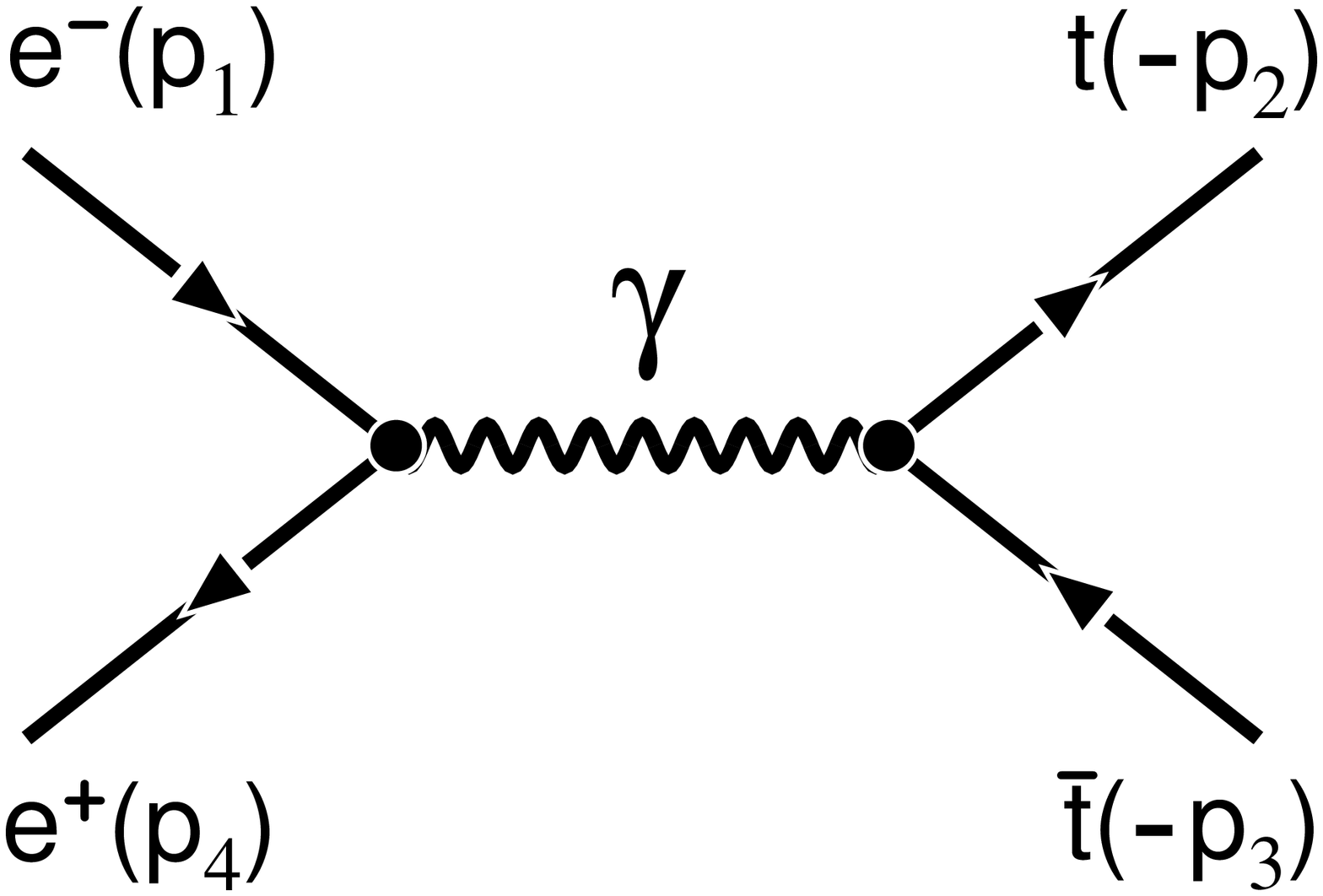}}
 \h{35}
\hspace*{0.4cm}
\mbox{\epsfysize=3.0cm\epsffile{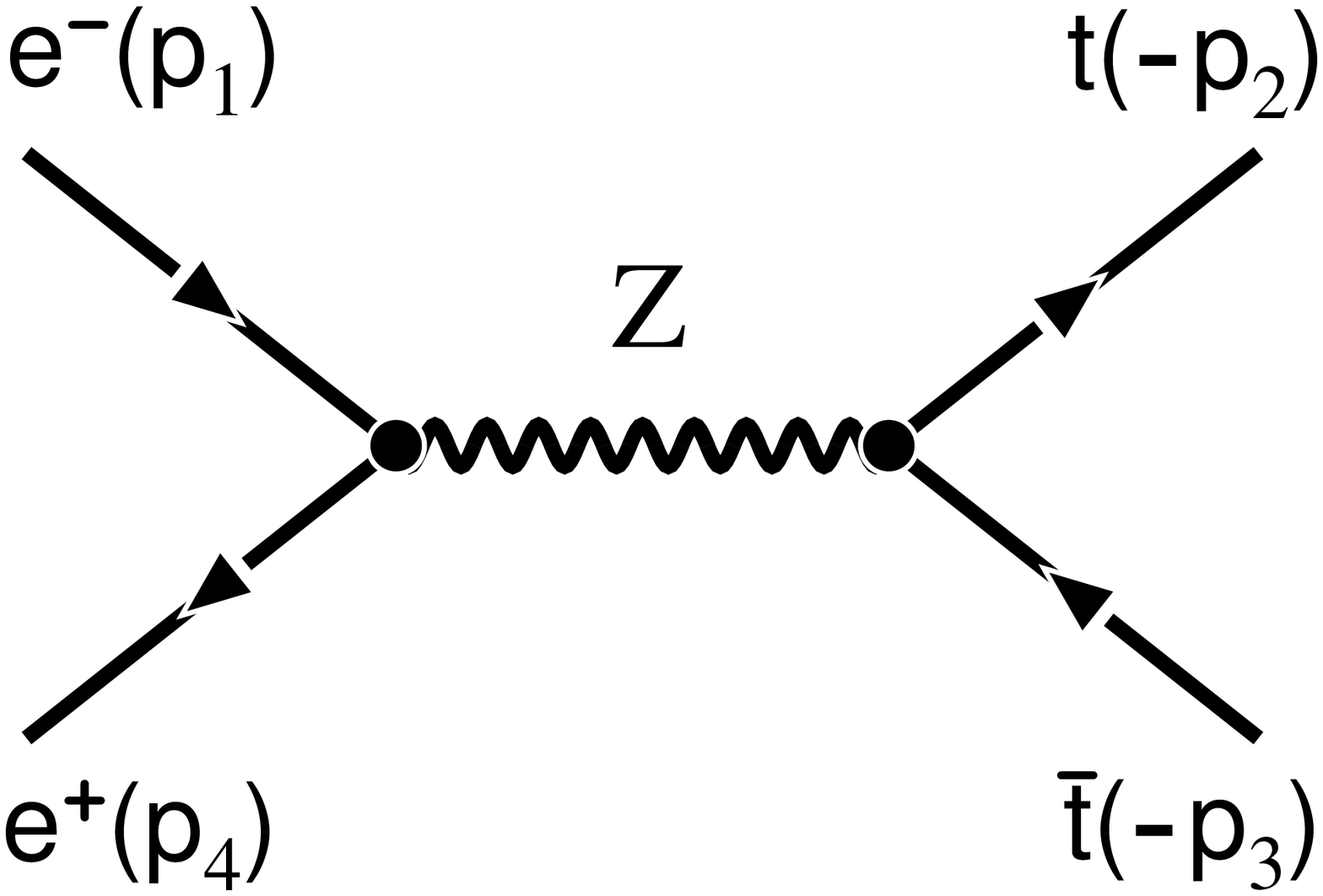}}
\vspace*{.6cm}
\caption{\label{feyntt0} {Feynman diagrams for the process \eett in
    Born approximation.}}
\end{center}
\end{figure}

In lowest  order perturbation  theory the process \eett can be
illustrated by the two Feynman diagrams of Fig.~\ref{feyntt0}.
For convenience we introduce the following abbreviations:
\begin{align}
  \label{eqn:momenta}
p_5 & = p_1 + p_2 = -p_3 - p_4, \h{10} p_5^2= t, 
\\   \label{eqn:momentas}
p_6 & = p_2 + p_3 = -p_1 - p_4, \h{10} p_6^2= s,  
\\   \label{eqn:momentau}
p_7 & = p_2 + p_4 = -p_1 - p_3, \h{10} p_7^2= u.
\end{align}

\ni
In the Feynman gauge the  matrix elements corresponding to
Fig.~\ref{feyntt0} are:

\begin{align}
{\cal M}_{\gamma} & = \frac{e^2}{s} ~ Q_e Q_t~  [\,  \vt
\gamma^{\mu} \uo]\, {\times}\, [\,  \ut \gamma_{\mu} \vo], 
  \\  
\label{eqn:bornZ}
{\cal M}_{Z} & = \frac{e^2} {s-\mz^2+i\mz \Gamma_Z} \,
[\,\vt \gamma^{\mu}\,(v_{e}-a_{e}\gamma_5)\,\uo] \,{\times} \,  
[\,\ut \gamma_{\mu}\,(v_{t}-a_{t}\gamma_5)\,\vo],
\end{align}  
with 
\begin{align}
 \label{eqn:cZ}
v_f&= \frac{T^3_f - 2 \, Q_f \, \swto}{ 2\, \sw\, \cw},
  \\
a_f&= \frac{T^3_f}{ 2\, \sw\, \cw} ,
\end{align}
where $T^3_f$ is the quantum number corresponding to the third component
of the weak isospin, $e\,Q_f$ the electromagnetic charge, and
$\theta_{\scriptscriptstyle w}$ the weak mixing angle.

 We parametrize  the radiative corrections by
means of form 
factors. 
Defining the following four matrix elements
\begin{align}
\label{eq:amplitude1}
{\cal M}_{1}^{ij} & = 
\left[ \vt \ga^{\mu}\, {\bf G}^i \, \uo\right] \, { \times}  \, 
\left[ \ut \ga_{\mu}\, {\bf G}^j \,\vo \right] 
, ~~~i,j=1,5,  
\end{align}
with ${\bf G}^1 = { 1}$ and ${\bf G}^5 = { \ga_{ 5}}$, 
the Born amplitude can be written in a compact form:
\begin{equation}
 \label{eqn:Mborn}
{\cal M}_{B} = {\cal M}_{\gamma} + {\cal M}_{Z } = 
{\sum}_{i,j=1,5}~{F_{1}^{ij,B}} \,\,  {\cal M}_{1}^{ij}  .
\end{equation}

The form factors are
\begin{align}
 \label{eqn:bornFF}
 {F_1^{11,B}} & =   v_e \, v_t \, 
\frac{e^2}{s - \mz^2 +i\mz\Gamma_Z}  ~+~ Q_e
 \, Q_t \, 
\frac{e^2}{s}{ ~~\equiv~~ F_1^{11,B,Z} + F_1^{11,B,\gamma}}, 
  \\
 {F_1^{15,B}} & =-  v_e \, a_t \, 
\frac{e^2}{s - \mz^2 +i\mz\Gamma_Z},   \\
 {F_1^{51,B}} & =-  v_t \, a_e \, 
\frac{e^2}{s - \mz^2 +i\mz\Gamma_Z},  \\
 {F_1^{55,B}} & =  a_e \, a_t \, 
\frac{e^2}{s - \mz^2 +i\mz\Gamma_Z}.
\end{align}

\ni
 Besides 
(\ref{eq:amplitude1}),  we find at 
one-loop level three further basic matrix 
element structures (in the limit of vanishing electron mass): 
\begin{equation}
 \label{eqn:Mborna}
{\cal M}_{ \rm{1loop}} = 
\sum_{a=1}^4 {\sum}_{i,j=1,5}
~ {F_{ a}^{ij,\rm{1loop}}} \,\,  {\cal M}_{a}^{ij},  
\end{equation}
with 
\begin{align}
\label{eq:amplitudes1to4}
{\cal M}_{1}^{ij} & = 
\ga^{\mu}\, {\bf G}^i \otimes \ga_{\mu}\, {\bf G}^j,
\\[4mm]
{\cal M}_{2}^{ij} & = \ps_2\, {\bf G}^i  \, \otimes \, \ps_4 \, {\bf G}^{ j}, 
\\[4mm]
{\cal M}_{3}^{ij} & = \ps_2\,{\bf G}^i \, \otimes \,  {\bf G}^{j},   
\\[4mm]
\label{eq:amplitudes2to4}
{\cal M}_{4}^{ij} & = \ga^{\mu}\, {\bf G}^{i} \, \otimes \, 
\ga_{\mu}\,\ps_4\, {\bf G}^{j}, 
\end{align}
and respectively sixteen scalar form factors $F_a^{ij}$ in total. 
An alternative notion uses the helicity structures,
\begin{align}
\label{eq:amplitudesLR}
{\cal M}_{1}^{LR} & = 
\ga^{\mu}\, {\bf L} \otimes \ga_{\mu}\, {\bf R}
\end{align}
etc., with ${\bf L},{\bf R} = (1\mp\gamma_5)/2$.
The interferences of these matrix elements with the Born amplitude have
  to be calculated. 
Only six of these interferences are independent, e.g. $ {\cal
  M}_{1}^{ij}$, ${{ \cal M}_3}^{11}$ and ${{ \cal M}_3}^{51}$,
\footnote{
We are grateful to D. Bardin and P. Christova for drawing our attention
to this simplification. 
} i.e. we have the following 10 equivalences :
{\allowdisplaybreaks
\begin{eqnarray}
 \label{eqn:16to6}
4~{{ \cal M}_2}^{11}&\leftrightarrow&
(T-U){{ \cal M}_1}^{11} + s~ {{ \cal M}_1}^{55} 
,  \\
4~{{ \cal M}_2}^{15}&\leftrightarrow& 
(T-U) {{ \cal M}_1}^{15} 
+ (s-4m_t^2)~ {{ \cal M}_1}^{51}
-~4{m_t} {{ \cal M}_3}^{51} 
,  \\
4~ {{ \cal M}_2}^{51}&\leftrightarrow& 
(T-U) {{ \cal M}_1}^{51} + {s}~ {{ \cal M}_1}^{15}
,  \\
4~ {{ \cal M}_2}^{55}&\leftrightarrow&
  (T-U) {{ \cal M}_1}^{55}+ (s-4{m_t^2})~ {{ \cal M}_1}^{11} 
-~4{m_t}~ {{ \cal M}_3}^{11} 
,  \\
{{ \cal M}_3}^{15} &\leftrightarrow& 0
,  \\
{{ \cal M}_3}^{55} &\leftrightarrow& 0
,  \\
 {{ \cal M}_4}^{55} \leftrightarrow -~{{ \cal M}_4}^{11} &\leftrightarrow& 
 {{ \cal M}_3}^{11} +{m_t}~ {{ \cal M}_1}^{11}
,  \\
{{ \cal M}_4}^{15} \leftrightarrow - {{ \cal M}_4}^{51} &\leftrightarrow&
{{ \cal M}_3}^{51}+m_t~{{ \cal M}_1}^{51}    \, .
 \label{eqn:16to6a}
\end{eqnarray}
}
{In the massless limit  ($m_t \to 0$) , only 
${\cal M}_1$ and ${\cal M}_2$ will 
contribute to the cross-section, and  it   can be
 expressed in terms of  the Born-like structures 
${\cal M}_1$ exclusively.}
We  introduce   
the variables
\begin{align}
 \label{eqn:defTUbeta}
T =   m_e^2 +  m_t^2 - t  &  \simeq   \frac{s}{2} \, (1-\beta\, \cos \, \theta) 
,\\
U =   m_e^2 +  m_t^2 -u &   \simeq    \frac{s}{2} \, (1+\beta \, \cos \, \theta)  
,\\
\beta_t = \beta &  = \sqrt{ 1- 4\, m_t^2/{s}}   \, .
\end{align}

Based on the relations (\ref{eqn:16to6}  to \ref{eqn:16to6a} )
  { the
  virtual corrections can be expressed  in terms of 
six
independent, modified, dimensionless} form factors $\widehat{F}_1^{ij }, \widehat{F}_3^{11},
\widehat{F}_3^{{51}}$:    
\begin{eqnarray}
  \label{eq:eqf61}
\widehat{F}_1^{11} &=&  \Bigl[ F_1^{11} 
+ \frac{1}{4}(u-t)  ~F_2^{11} 
 -   \frac{1}{4}(u  +   t  +  2m_t^2)   ~F_2^{55} 
+ m_t~  (F_4^{55}-F_4^{11} ) \Bigr],
\\
\widehat{F}_{ 1}^{ 15} &=& \Bigl[F_{ 1}^{ 15} 
- \frac{1}{4} (u+t-2m_t^2) ~F_{ 2}^{ 51} 
+ \frac{1}{4} (u-t) ~F_{ 2}^{ 15} \Bigr],
\\
\widehat{F}_{ 1}^{ 51} &=& \Bigl[F_{ 1}^{ 51} 
+ \frac{1}{4} (u-t) ~F_{ 2}^{ 51}  
- \frac{1}{4} (u+t+2m_t^2)~F_{ 2}^{ 15}  
+m_t~(F_{ 4}^{ 15}-F_{ 4}^{ 51})\Bigr],
\\
\widehat{F}_1^{55} &=&  \Bigl[F_1^{55} 
- \frac{1}{4} (u+t-2m_t^2   ) ~F_2^{11}
 +  \frac{1}{4} (u-t)~F_2^{55} \Bigr],
\\
  \label{eq:eqf63}
\widehat{F}_3^{11}&=& \Bigl[F_3^{11}-F_4^{11}+F_4^{55}-m_t ~F_2^{55}\Bigr],
\\
\widehat{F}_{ 3}^{ 51} &=& \Bigl[{F}_{ 3}^{ 51} +
F_{ 4}^{
15}-F_{ 4}^{ 51}-m_t ~F_{ 2}^{ 15} \Bigr]  \, .
\end{eqnarray}
The resulting  cross-section formula is:
\begin{eqnarray}
  \label{eq:sigma6f}
  \frac{d\sigma}{d\cos\theta} &=& 
\frac{\pi \alpha^2}{2s} 
~c_t~ 
\beta
~2 \Re e \Bigl[ 
  (u^2+t^2+2m_t^2s)
\left(\bar{F}_1^{11} \bar{F}_1^{11,B*} +\bar{F}_1^{51} \bar{F}_1^{51,B*} \right)
\nonumber\\
&&+~
  (u^2+t^2-2m_t^2s)
\left(\bar{F}_1^{15} \bar{F}_1^{15,B*} +\bar{F}_1^{55} \bar{F}_1^{55,B*} \right)
\nonumber\\
&&
+~  (u^2-t^2)
\left( \bar{F}_1^{55} \bar{F}_1^{11,B*}+ \bar{F}_1^{15} \bar{F}_1^{15,B*}
  + \bar{F}_1^{51} \bar{F}_1^{51,B*}    + \bar{F}_1^{11} \bar{F}_1^{55,B*} \right)
\nonumber\\
&&
+~ 2m_t(tu-m_t^4) \left( \bar{F}_3^{11}  \bar{F}_1^{11,B*} 
                       +\bar{F}_3^{51}  \bar{F}_1^{51,B*} \right)
\Bigr],
\end{eqnarray}
 where the dimensionless form factors are 
\ba
\label{starbo} 
\bar{F}_1^{ij,B*}  &=& \frac{s}{e^2}~F_1^{ij,B*} ,
\\
  \label{eq:sigma6h}
 \bar{F}_{ a}^{ ij} &=& 
\frac{s}{e^2}~\left[
\frac{1}{2} \delta_{a,1} 
{F}_{ 1}^{ ij,B} +  
{ \frac{1}{16\pi^2}}~
\widehat {F}_{ a}^{ ij, \rm{1loop}}\right].
\ea
\ni 
 and  $c_t  = 3$, $\alpha=e^2/4\pi$. 
 The $\bar{F}_a^{ij}$ are defined so that  
double counting for the Born contributions ${F}_{1}^{ij,B}$ is
avoided.
The factor $1/(16\pi^2)$ is conventional. 

In the numerical program,  helicity form
factors are calculated as well.   They are defined as follows:
\allowdisplaybreaks
\begin{eqnarray}
\label{eqn:15toLR}
         F_i^{LL}& = & \frac{1}{4}\left[ F_i^{11}-F_i^{15}-F_i^{51}+F_i^{55}\right], 
\\  
         F_i^{LR}& = &  \frac{1}{4}\left[F_i^{11}+F_i^{15}-F_i^{51}-F_i^{55} \right],
\\ 
         F_i^{RL}& = &  \frac{1}{4}\left[F_i^{11}-F_i^{15}+F_i^{51}-F_i^{55} \right],
\\ 
         F_i^{RR}& = &  \frac{1}{4}\left[F_i^{11}+F_i^{15}+F_i^{51}+F_i^{55}\right]
         ,~~~~i=1\ldots4 .
\end{eqnarray}
At the end of this introductory section, we would like to give the
relation of our form factors to those used in the literature for pair
production of massless fermions.
In that case, only the four form factors $\widehat F_{1}^{ij}$ contribute.
They have to replace, in the massless limit, the form factors $\rho$
and $\kappa_f$, 
which are conventionally used to renormalize the muon decay
constant and the weak mixing angle and are precisely defined in 
\cite{Bardin:1989di,Bardin:1999yd}.
We rewrite the matrix element ${\cal M}_1$ in such a way that it gives
exactly the Born $Z$ amplitude (\ref{eqn:bornZ}) when the four 
form factors $\rho_{et},\kappa_e,\kappa_t,\kappa_{et}$ are set equal
to 1: 
\begin{align}
\label{eq:ampl1a}
{\cal M}_{1} & = 
\sum_{i,j=1,5}  \widehat {F}  _1^{ij} {\cal M}_{1}^{ij} \equiv  
 \sum_{i,j=L,R} \widehat {F} _1^{ij} {\cal M}_{1}^{ij}
\nonumber \\
&=
\frac{4 \, e^2 \, {a}_e \, {a}_t} {s-\mz^2+i\mz \Gamma_Z} 
~\rho_{et} \, 
\Biggl[
~\left(\gamma^{\mu}{\bf L} \otimes \gamma_{\mu}{\bf L}\right)
- |Q_{ e}|  \swto  \kappa_{ e} 
~\left(\gamma^{\mu}\otimes \gamma_{\mu}{\bf L}\right)
\nonumber \\
& \hspace{2cm} - |Q_{ t}|  \swto  \kappa_{ t} 
~\left(\gamma^{\mu}{\bf L} \otimes \gamma_{\mu}\right)
+ |Q_t \, Q_e|  \swfor  \kappa_{et} 
~\left(\gamma^{\mu}\otimes \gamma_{\mu}\right)
\Biggr].
\end{align}
From here, it is easy to derive the relations between the form
factors $\widehat {F}_{ 1}^{ ij}$ in an (L,R) or (1,5) basis and $\rho_{et},
\kappa_e,\kappa_t,\kappa_{et}$,
respectively: 
\begin{align}
\label{eqn:lrrk}
\widehat {F}_{ 1}^{ LL} & = \frac{4\, \,e^2 \,{a}_e \,{a}_t}
{s-\mz^2+i\mz \Gamma_Z}
~\rho_{et}
\left( 1- |Q_e|  \swto   \kappa_e - |Q_t|  \swto\, 
\kappa_t+ |Q_t \,  Q_e|  \swfor  \kappa_{et} \right),
 \\
\widehat {F}_{ 1}^{ LR} & = \frac{4\,\, e^2\, {a}_e\, {a}_t}
{s-\mz^2+i\mz \Gamma_Z}
~\rho_{et}
\left(-|Q_{ t}|  \swto \,  \kappa_{ t} + |Q_t \,
Q_e|  \swfor   \kappa_{et} \right),
 \\
\widehat {F}_{ 1}^{ RL} & = \frac{4\,\, e^2\, {a}_e\, {a}_t}
{s-\mz^2+i\mz \Gamma_Z}
~\rho_{et}
\left(-|Q_{ e}|  \swto\,   \kappa_{ e} +  |Q_t \, Q_e|
 \swfor   \kappa_{et} \right),
 \\
\widehat {F}_{ 1}^{ RR} & = \frac{4\,\, e^2\, {a}_e\, {a}_t}
{s-\mz^2+i\mz \Gamma_Z} 
~\rho_{et}
\left(|Q_t \,  Q_e|  \swfor\,   \kappa_{et} \right).
\end{align}
Three process-dependent  effective weak
mixing angles $ \swto^{\rm eff}$ and the 
weak coupling strength $\kappa^{\rm eff}$ are obtained by inverting these
relations: 

\allowdisplaybreaks
\begin{eqnarray}
\label{eqn:rklr}
\kappa^{\rm eff} = \rho_{et} 
&=& 
\frac{s-\mz^2+i\mz  \Gamma_Z}{4\,\, e^2\, {a}_e\, {a}_t } 
~\left(\widehat F_1^{LL}-\widehat F_1^{LR}-\widehat F_1^{RL}+\widehat F_1^{RR} \right) ,
 \\
{ \swto}^{{\rm eff}, \, e} = \kappa_e ~\swto
&=& - \frac{1}{|Q_e|} 
~\frac{\left(\widehat F_1^{RR} - \widehat F_{ 1}^{ RL}\right)}
{\left(\widehat F_1^{LL}-\widehat F_1^{LR}-\widehat F_1^{RL}+\widehat F_1^{RR} \right)},
 \\
{\swto}^{{\rm eff}, \, t} = 
\kappa_t ~\swto  &=& - \frac{1}{|Q_t|}
~\frac{\left(\widehat F_1^{RR}-\widehat F_{ 1}^{ LR}\right)}
{\left(\widehat F_1^{LL}-\widehat F_1^{LR}-\widehat F_1^{RL}+\widehat F_1^{RR} \right)},
 \\
({ \swto}^{{\rm eff},  e\,t})^2 = 
\kappa_{et} ~\swfor  &=&  \frac{1}{|Q_{e}Q_t|}
~\frac{\widehat F_1^{RR}}
{\left(\widehat F_1^{LL}-\widehat F_1^{LR}-\widehat F_1^{RL}+\widehat F_1^{RR}\right)} .
\end{eqnarray}
For the simplest approximations with factorizing, universal weak
corrections, the $\kappa_e$, $\kappa_t$, and {$\sqrt{\kappa_{et}}$}
become equal, real 
and constant (independent of process and kinematics); for more details
see for instance the discussion of the weak corrections in
\cite{Bardin:1997xq,Kobel:2000aw} and references therein.
There, also the important influence of higher order corrections is
considered. 

\section{Virtual Corrections \label{sec-virtual}}
The virtual corrections come from self-energy insertions, vertex and
box diagrams, and from renormalization.
A complete list of the contributing diagrams may be found in
\cite{Fleischer:2002rn}.   
By means of the package 
DIANA \cite{Tentyukov:1999is,Tentyukov:1999yq,Fleischer:2000zr} we
generated
useful graphical presentations of the diagrams and the input for subsequent FORM
\cite{FORM,Vermaseren:2000nd} manipulations. 
With the DIANA output  (FORM input), we performed two independent calculations of
the virtual form factors, both using the 't Hooft--Feynman gauge.

For the final numerical evaluations we used two Fortran packages: 
{\tt FF}
\cite{vanOldenborgh:1991yc} and {\tt LoopTools} \cite{Hahn:1998yk}. 
Both have been taken from the corresponding homepages, and {\tt LoopTools}
was slightly adapted: one {infrared $C_0$ was added and the $DB_1$
was used only for photon mass $\lambda = 0$}.

 In the package 
{\tt FF}, the Passarino--Veltman tensor decomposition of the
amplitudes \cite{Passarino:1979jh} is defined in terms of
the external momenta of the diagrams, while in  {\tt LoopTools}
  this
decomposition is performed in terms of internal momenta and the latter
are later  expressed in terms of  the external ones.
The resulting linear relation between the corresponding form factors is
given in Appendix \ref{app-extint}.
Both the ultraviolet (UV) and the infrared (IR) divergences are
treated by dimensional regularization, introducing the dimension 
$d=4-2 \epsilon$ and parametrizing the infinities as poles in 
$\epsilon$.
The 
 UV divergences 
have to be eliminated by renormalization
on the amplitude level, while the 
 IR  ones can only be eliminated on the
cross-section level by including the emission of soft photons. 
For the IR divergences we have alternatively  introduced 
a finite photon mass, as is foreseen in {\tt FF}, yielding a logarithmic
singularity in this mass. 
Agreement to high precision was achieved for the two approaches. 

Because the calculation of one-loop corrections 
 for 
{two-fermion production} is well known, we do
  not present a detailed prescription
of the calculations.
{We perform the  renormalization  closely following 
\cite{Fleischer:1981ub}}. 
 On  the other hand, we want to fix some cornerstones and 
sketch the renormalization 
and 
show the 
UV-divergent parts of all the contributing diagrams,
 such that their  cancellations  can be deduced.
The treatment of the IR divergences will be discussed in 
more detail because of the interplay with real photonic corrections.
 Finally, concerning the finite parts, we refer to the Fortran program {\tt
  topfit}.
We only mention that we did not perform a complete reduction of the
various scalar functions 
to $A_0$, $B_0$, $C_0$, and $D_0$,
since this is not needed for a purely numerical evaluation. 


\subsection{The self-energy diagrams \label{sec-selfe}}
We have {to renormalize} the UV singularities of the self-energies
of the photon and the $Z$ boson, and also that of their mixing. 
Since the counter terms from wave-function
and parameter renormalization must exhibit Born-like structures, it is clear
from the very beginning that a cancellation of UV divergences can only
occur in terms of single propagator poles. 
The double poles, which originally occur in the self-energy
diagrams, are cancelled by mass renormalization. 
This we want to stress  for  the 
following, by allowing only single poles in the self-energy
contributions.
Thus 
the 
photon self-energy and the $Z$ self-energy
diagrams take  the form 
\begin{eqnarray}
\label{ren-10}
S_{\gamma}&=& 
\frac{{\Sigma}_{\gamma \gamma}}{s} ~~  {\cal M}_{\gamma}
,
\\
\label{ren-11}
S_{Z}&=&
  \frac{{\Sigma}_{Z Z}}{s - \mz^2}  
~~{\cal M}_{Z }
.
\end{eqnarray}
In the $\gamma$--$Z$ mixing diagrams, a partial fraction
decomposition of the product $(1/s) 1/(s - \mz^2)$ is
performed, but no subtraction: 
\begin{eqnarray}
\label{ren-12}
S_{\gamma Z}&=&  
- e^2 \, Q_e~
         {\Sigma}_{\gamma Z}~  \frac{1}{\mz^2}
         \left(\frac{1}{s - \mz^2}-\frac{1}{s}\right)  
{\gamma}_{\mu}  \times
         {\gamma}_{\mu} \left( v_t - a_t {\gamma}_5 \right) , 
\\
\label{ren-13}
S_{Z \gamma}&=&
  -e^2 \, 
 Q_t  {\Sigma}_{\gamma Z}~    \frac{1}{\mz^2}
 (\frac{1}{s - \mz^2}-\frac{1}{s})
{\gamma}_{\mu}  ( v_e - a_e {\gamma}_5 )   \times 
         {\gamma}_{\mu}.
\end{eqnarray}
We give
the UV-divergent parts of the  renormalized 
self-energies (for definitions, see Appendix \ref{app-re}): 
\allowdisplaybreaks
\begin{eqnarray}
\label{ren-15}
{\Sigma}_{\gamma \gamma}^{\rm  ren,  UV} &=&  e^2 \, s
\left( - \frac{23}{3} \right) ~\frac{1}{\epsilon} ,
\\
\label{ren-16}
{\Sigma}_{Z\gamma}^{\rm  ren,  UV}= {\Sigma}_{\gamma Z}^{\rm  ren,  UV} &=&  \frac{e^2}{2\,
\sw \cw} \left( \frac{46}{3} \cwto~ s -4 \mw^2
- \frac{41}{3} s \right)          ~\frac{1}{\epsilon} ,
\\
\label{ren-17}
{\Sigma}_{Z Z}^{\rm  ren,   UV}&=& e^2 \,  (s - \mz^2) 
 \left( \frac{23}{3} + \frac{6} {\swto} - 
\frac{41}{6 ~ \cwto \swto} \right)
                     ~\frac{1}{\epsilon} . 
\end{eqnarray}
Three families of fermions are assumed.
The UV-divergent terms of the self-energies are independent of the fermion masses, but
this is, of course,  not true for the finite contributions.

The form factors are easily deduced from the above representations.
The photon self-energy, for instance, contributes to $F_{1}^{11}$ only:
\begin{eqnarray}
\label{ren-14}
\widehat{F}_{1}^{11,\gamma \gamma} = 
F_1^{11,B,\gamma \gamma} ~~\frac{{\Sigma}_{\gamma \gamma}^{\rm ren}}{s}
\equiv
Q_e \, Q_t \, \frac{e^2}{s^2}~~{{\Sigma}_{\gamma \gamma}^{\rm ren}} .
\end{eqnarray}


\subsection{The vertex diagrams \label{subsec-vertices}}
From the initial-state vertex corrections, form factors  
 $F_1^{ij,V}, V=\gamma, Z$, 
arise, and from the final vertices  
 $F_1^{ij,V}$ and $F_3^{ij,V}$,  
the latter being proportional to $m_t$.
There are UV divergences from the vertex diagrams in $F_1^{ij}$: 
 again  only Born-like amplitudes are UV-divergent.

The divergent parts of vertices with a photon or $Z$ boson in the
$s$-channel, correspondingly,  are:\footnote {To compactify the
following formula we introduce the abbreviations: $\sw = s_W$ and $
\cw = c_W$.}
\begin{eqnarray}
\label{ren-1}
V_{\gamma}^{\rm UV} &=&
\left( \frac{e^2}{2s_W}\right)^2~\frac{1}{s} ~ \frac{1}{\epsilon}
\left[
f^{11,\gamma}_{1}          {\gamma}_{\mu} \otimes {{\gamma}^{\mu}} 
+ f_{1}^{ 15 ,\gamma} {\gamma}_{\mu} \otimes {{\gamma}^{\mu}} {\gamma}_5 
+ f_{1}^{ 51 ,\gamma} {\gamma}_{\mu} {\gamma}_5 \otimes {{\gamma}^{\mu}} 
\right],
\\
\label{ren-2}
V_Z^{\rm UV} &=& 
 \left( \frac{e^2}{2s_W^2c_W}\right)^2~\frac{1}{s-\mz^2} ~ \frac{1}{\epsilon}
\left[
f^{11,Z}_{1}             {\gamma}_{\mu} \otimes {{\gamma}^{\mu}} 
+ f_1^{ 15 ,Z} {\gamma}_{\mu} \otimes {{\gamma}^{\mu}} {\gamma}_5 
+ f_1^{ 51 ,Z} {\gamma}_{\mu} {\gamma}_5 \otimes {{\gamma}^{\mu}} 
 + f_1^{55,Z}            {\gamma}_{\mu} {\gamma}_5 \otimes {{\gamma}^{\mu}} {\gamma}_5 
\right] .
\nonumber \\
\end{eqnarray}
The explicit expressions from the initial and final photonic vertices are:
\begin{eqnarray}
\label{ren-3}
f_1^{11,\gamma} &=&   \left(  - \frac{17}{27} \frac{1}{c_W^2}\,   - \frac{64}{27} -
\frac{ m_t^2}{\mw^2} - \frac{1}{3} \, 
\frac{m_b^2}{\mw^2} \right)_{\rm fin} 
-  \left(  \frac{5}{3} \frac{1}{c_W^2}  + \frac{2}{3}
\right)_{\rm ini}
,
\\
\label{ren-5}
f_1^{ 15  ,\gamma} &=&  \left( \frac{32}{9} - \frac{5}{9} \frac{1}{c_W^{2}} -
\frac{1}{3}\frac{m_t^2}{\mw^2} 
          + \frac{1}{3}\frac{m_b^2}{\mw^2}   \right)_{\rm fin}
,
\\ 
\label{ren-4}
f_1^{ 51 ,\gamma} &=&    \left(  \frac{10}{3} - \frac{1}{c_W^{2}}
\right)_{\rm ini}.
\end{eqnarray}
 For the $Z$ boson in the $s$-channel we only give the sum of the
initial- and final-state fermion vertices:
\begin{eqnarray}
\label{ren-6}
f^{11,Z}_{1} &=& \frac{973}{216} - \frac{25}{18}\frac{1}{c_W^{2}}
          - \frac{9}{16}\frac{m_t^2}{\mw^2} - \frac{1}{16}\frac{m_b^2}{\mw^2}
          -c_W^2 \frac{m_t^2}{\mz^2} - \frac{1}{3} c_W^2 \frac{m_b^2}{\mz^2}
          - \frac{157}{108}c_W^2 - \frac{82}{27}c_W^4 
          + \frac{3}{2}\frac{m_t^2}{\mz^2} +
          \frac{1}{3}\frac{m_b^2}{\mz^2}
,
\nll
\\  
\label{ren-8}
f^{15,Z}_{1} &=&  \frac{21}{8} -\frac{1}{c_W^{2}} - \frac{7}{16}\frac{m_t^2}{\mw^2}
          + \frac{1}{16}\frac{m_b^2}{\mw^2} - \frac{1}{3}c_W^2
          \frac{m_t^2}{\mz^2} 
        + \frac{1}{3} c_W^2 \frac{m_b^2}{\mz^2}- \frac{137}{36}c_W^2 
         + \frac{32}{9}c_W^4 + \frac{5}{6}\frac{m_t^2}{\mz^2}
          - \frac{1}{3}\frac{m_b^2}{\mz^2} 
,
\nll
\\
\label{ren-7}
f^{51,Z}_{1} &=&  \frac{665}{216} - \frac{95}{108}\frac{1}{c_W^{2}}
          - \frac{3}{16}\frac{m_t^2}{\mw^2} - \frac{1}{48} \frac{m_b^2}{\mw^2}
          - \frac{449}{108} c_W^2 + \frac{10}{3}c_W^4
          + \frac{1}{4}\frac{m_t^2}{\mz^2} +
          \frac{1}{12}\frac{m_b^2}{\mz^2}  
,
\\ 
\label{ren-9}
f^{55,Z}_{1} &=& \frac{97}{72} - \frac{7}{12}\frac{1}{c_W^{2}} -
\frac{7}{48}\frac{m_t^2}{\mw^2} 
        + \frac{1}{48}\frac{m_b^2}{\mw^2} - \frac{77}{36} c_W^2
        + \frac{1}{12}\frac{m_t^2}{\mz^2} -
        \frac{1}{12}\frac{m_b^2}{\mz^2}.
\end{eqnarray}

The resulting form factors can be extracted, 
e.g.
\begin{eqnarray}
  \label{eq-vertg}
\widehat{F}_1^{ij,\gamma, \rm UV} 
&=&   (e^4)/(4s_W^2 s \epsilon) ~f_1^{ij,\gamma}.
\end{eqnarray}

\subsection{The box diagrams \label{subsec-boxes}}
$ZZ$, $Z \gamma$ and $\gamma Z$  box diagrams contribute to
all 
form factors  
$F_1^{ij}$ to $F_4^{ij}$ introduced in (\ref{eqn:Mborna}) to
(\ref{eq:amplitudes2to4}), while the $WW$ box diagram contributes only to 
$F_1^{ij}$.
The pure photonic box diagrams contribute only to $F_a^{11}$ and $F_a^{55}$.
Simple power counting shows that
there are no UV divergences 
from the boxes.
The IR divergences will be discussed in Section \ref{sec-infrared} and
Appendix \ref{chap:IR}.

\subsection{The counter-term contributions \label{sec-countert}}
Finally we have to take into account the contribution from the
counter terms of Appendix \ref{app-re}, where we also
introduce some of the notation to be used in the following.
 With these the  photon exchange becomes: 
\begin{eqnarray}
\label{ren-16a}
C_{\gamma}&=&\left[ 
         {\gamma}_{\mu}  \otimes
         {\gamma}_{\mu}~  ( z_{a,t} + z_{b,t} {\gamma}_5 )  +
         {\gamma}_{\mu}~  ( z_{a,e} + z_{b,e} {\gamma}_5 )   \otimes
         {\gamma}_{\mu}  +  
   {\gamma}_{\mu} \otimes {{\gamma}^{\mu}}~ 2 \frac{\delta e}{e}  \right] 
         Q_e Q_t  \frac{e^2}{ s} .
\end{eqnarray}
Analogously,  for the  $Z$ exchange 
\begin{eqnarray}
\label{ren-17a}
C_{Z}&=&\Biggl\{ 
        {\gamma}_{\mu}  ( v_e - a_e {\gamma}_5 )  \otimes 
         {\gamma}_{\mu}  ( v_t - a_t {\gamma}_5 ) 
         ( z_{a,t} + z_{b,t} {\gamma}_5 ) 
 +  ~       {\gamma}_{\mu}  ( v_e - a_e {\gamma}_5 )
  ( z_{a,e} + z_{b,e} {\gamma}_5 ) \otimes
          {\gamma}_{\mu}  ( v_t - a_t {\gamma}_5 ) 
\nonumber 
\\
&&-~       {\gamma}_{\mu} \otimes
         {\gamma}_{\mu}\,   ( v_t - a_t {\gamma}_5 )  ~
\frac{Q_e}{ s_W \, c_W }  {\delta s_W^2} 
-~
       {\gamma}_{\mu}  ( v_e - a_e {\gamma}_5 )  \otimes
{\gamma}_{\mu}  \frac{ Q_t}{ s_W \, c_W  }  {\delta
s_W^2}
 \nonumber \\
&&+~
       {\gamma}_{\mu}   ( v_e - a_e {\gamma}_5 )  \otimes 
         {\gamma}_{\mu}  ( v_t - a_t {\gamma}_5 ) 
         \left[2 \frac{\delta e}{e} + 
\left( \frac{1}{c_W^{2}} - \frac{1}{s_W^{2}}\right) \delta
         s_W^2\right] \Biggr\}  
         \frac{e^2}{s - \mz^2} .
\end{eqnarray}
It is again easy to collect from the above expressions the
corresponding contributions to the form factors $F_1^{11}$ to $F_1^{55}$. 
The contributions to, say,  $F_1^{11}$ from the counter terms are:
\begin{eqnarray}
\label{ren-17b}
\widehat{F}_1^{11,ct} &=& \left[2 \frac{\delta e}{e} + z_{a,t} +
  z_{a,e} \right] F_1^{11,B,\gamma} 
+~\left[ 2 \frac{\delta e}{e} + 
 z_{a,t} + z_{a,e} +
\left( \frac{1}{c_W^{2}} - \frac{1}{s_W^{2}}\right) \delta s_W^2
\right] F_1^{11,B,Z}
\nll && 
-~(v_tQ_e+v_eQ_t)~\frac{\delta s_W^2}{s_W\,c_W}~\frac{e^2}{s - \mz^2} .
\end{eqnarray}
The resulting $1/\epsilon$ terms may be read off in Appendix \ref{app-re}.

\bigskip

The sum of all the ${1}/{\epsilon}$ terms listed in the foregoing subsections  
for the form factors $F_1^{ij}$, $i,j = 1,\ldots,4$, has been shown to
finally vanish separately for the photon and the  $Z$  pole of the
s-channel propagator:
\begin{eqnarray}
\label{ren-17c}
\widehat{F}_1^{ij,\rm UV} &=& \left[ 
\widehat{F}_1^{ij,\gamma\gamma} + \widehat{F}_1^{ij,\gamma Z} +
\widehat{F}_1^{ij,Z\gamma} + \widehat{F}_1^{ij,ZZ} +
\widehat{F}_1^{ij,\gamma} + \widehat{F}_1^{ij,Z} + \widehat{F}_1^{ij,ct}
\right]_{\rm UV} 
~~=~~0. 
\end{eqnarray}


\subsection{Infrared divergences \label{sec-infrared}}
\bigskip

In {\tt topfit}, we have {incorporated} two weak libraries.
One uses the package {\tt LoopTools} \cite{Hahn:1998yk}, and the other
one the package {\tt FF} \cite{vanOldenborgh:1991yc}.
With these two options, we have a variety of internal cross
checks at our disposal.

The photonic virtual corrections contain infrared divergences.
They appear  as  singular  behaviour of classes 
of one-loop functions.
One may follow several strategies to handle them in a numerical
calculation.
The simplest one is to blindly give the task to the library for
numerical calculation of the one-loop functions and then control the
infrared stability numerically in the Fortran program.
Both packages allow for this approach;  {\tt LoopTools} with
dimensional regularization or with a finite photon mass, while 
{\tt FF} treats loop functions with finite photon mass only.

In addition, we checked the IR stability in two ways explicitly.
In the library based on {\tt FF}, 
 we simply took a small but finite photon mass and directly
applied {\tt FF} without simplifying any tensor functions. 
Several analytic checks were also performed.
In the other one, we isolated in all the scalar functions the
IR divergence explicitly and
the cancellations   with the divergences from 
bremsstrahlung corrections (see Section \ref{softint}) were  controlled both
analytically and  numerically.   

In Appendix \ref{chap:IR}, we fix the notation and discuss the basics
of the treatment of IR divergences.
In the rest of this section, we simply give a list of the divergent
parts of the form factors:

From the renormalization of the fermion self-energies  (wave function
renormalization factors) :
\begin{eqnarray}
\label{zfIR}
\widehat{F}_1^{ij,Z_f,\rm IR} &=& 
- 4\, e^2 Q_f^2
m_f^2 ~ DB_1(m_f^2;m_f^2,0) ~ 
{F}_1^{ij,B} .
\end{eqnarray}

From the vertex corrections (index $f = e,t$):
\begin{eqnarray}
\label{vertIR}
\widehat{F}_1^{ij{,{V_f},}\rm IR} &=& - 2\, e^2 Q_f^2
(s-2m_f^2)  ~ C_0(m_f^2,s,m_f^2;0,m_f^2,m_f^2) ~ {F}_1^{ij,B} .
\end{eqnarray}

These form factors combine in the cross-section with the initial- and
final-state soft photon corrections to an infrared-finite contribution.  
For instance, the pure photonic parts contribute only to $\bar{F}_1^{11}$.
The resulting IR-divergent cross-section contribution in
(\ref{eq:sigma6f}),
\begin{eqnarray}
\label{sigIR}
  \frac{d\sigma^{ f,\rm IR }}{d\cos\theta} &=&
  \frac{d\sigma^B}{d\cos\theta} 
~~\frac{\alpha}{\pi} ~~Q_f^2~~\delta_f^{\rm IR},
\end{eqnarray}
with
\begin{eqnarray}
\label{delIR}
\delta_f^{\rm IR} = 2 ~\ln\frac{m_f}{\lambda}~ \left( 1 + 
\frac{s-2m_f^2}{s\beta_f}~\ln\frac{1-\beta_f}{1+\beta_f} \right),
\end{eqnarray}
is compensated with (\ref{delsini}) and
 (\ref{delsofin}). 

A little more involved are the box diagram contributions.
As a typical example, we show the photonic box parts.
The direct box gives:
\begin{eqnarray}
\label{dbxgIR}
{F}_1^{11,d \gamma,\rm IR} = - {F}_1^{55,d \gamma,\rm IR} &=&
(t-u-s)~{\cal G}_d, 
\\
{F}_2^{11,d \gamma,\rm IR} = {F}_2^{55,d \gamma,\rm IR} &=& -4~{\cal G}_d,
\\
{F}_4^{55,d \gamma,\rm IR} &=& -4m_t~{\cal G}_d,
\end{eqnarray}
with 
\begin{eqnarray}
\label{calG}
{\cal G}_d&=& 
-e^2~Q_eQ_t~~  C_0(m_e^2,t,m_t^2;0,m_e^2,m_t^2)~~{F}_1^{11,B,\gamma}.
\end{eqnarray}
From these expressions, the form factors 
 (\ref{eq:eqf61}) to (\ref{eq:eqf63})  
get  contributions:
\begin{eqnarray}
\label{cbxgIRw}
\widehat{F}_1^{11,d \gamma,\rm IR} &=&
+4 T~~e^2~Q_eQ_t~~  C_0(m_e^2,t,m_t^2;0,m_e^2,m_t^2)
~~{F}_1^{11,B,\gamma},
\end{eqnarray}
and all the others vanish.
Analogously, from the crossed photonic box diagram:
\begin{eqnarray}
\label{dbxgIR2}
{F}_1^{11,c \gamma,\rm IR} = + {F}_1^{155,c \gamma,\rm IR} &=&
-2(u-m_t^2)~{\cal G}_c, 
\\
{F}_2^{11,c \gamma,\rm IR} =- {F}_2^{55,c \gamma,\rm IR} &=& -4~{\cal G}_c,
\\
{F}_4^{55,c \gamma,\rm IR} &=& +4m_t~{\cal G}_c,
\end{eqnarray}
with 
\begin{eqnarray}
\label{calGc}
{\cal G}_c&=& 
-e^2~Q_eQ_t~~  C_0(m_e^2,u,m_t^2;0,m_e^2,m_t^2)~~{F}_1^{11,B,\gamma}.
\end{eqnarray}
From these expressions, the {form factors} 
 (\ref{eq:eqf61}) to (\ref{eq:eqf63})  
get  contributions:
\begin{eqnarray}
\label{cbxgIRw2}
\widehat{F}_1^{11,d \gamma,\rm IR} &=&
-4 U~~e^2~Q_eQ_t~~  C_0(m_e^2,u,m_t^2;0,m_e^2,m_t^2)
~~{F}_1^{11,B,\gamma},
\end{eqnarray}
and again all the others vanish.

The resulting cross-section {contributions become}
\begin{eqnarray}
\label{sigIRb}
  \frac{d\sigma^{ \gamma,\rm IR}}{d\cos\theta} &=&
  \frac{d\sigma^B}{d\cos\theta} 
~~4~\frac{\alpha}{\pi} ~~Q_eQ_t~~\ln\frac{1}{\lambda}
\left(\ln\frac{m_em_t}{T}-\ln\frac{m_em_t}{U}\right)  , 
\end{eqnarray}
and are compensated with (\ref{delsoint}). 

In Appendix \ref{chap:IR} we show the relevant scalar functions explicitly.


\section{Real Photonic Radiative Corrections}
\subsection{The three-particle phase space}
\label{sec-reaction}
The reaction 
\bq
e^+(p_4) ~+~ e^-(p_1)\rightarrow t(q_2)~+~\bar t(q_3)~+~\gamma(p)
\label{reaction5}
\eq
with
\ba
d\sigma &=& 
\frac{1}{2s\beta_0}
  |{\cal M}|^2\cdot (2\pi)^4\delta^4(p_1+p_4-q_2-q_3-p)
\frac{\mbox{d}^3\vec q_2}{(2\pi)^32E_t}
\frac{\mbox{d}^3\vec q_3}{(2\pi)^32E_{\bar t}}
\frac{\mbox{d}^3\vec p}{(2\pi)^32E_\gamma}
\label{crossint}
\ea
 is the one which in reality always takes place,
even if for soft photons the `elastic' Born cross-section
can be a good approximation without taking into account
the radiated photons. 
Here we introduce the final-state phase-space parametrization:
for convenience of notation, the top physical momenta 
$q_2=-p_2,q_3=-p_3$ are used here.
We will not neglect the electron mass systematically,
$p_1^2=p_4^2=m_e^2$, and
$\beta_0=\sqrt{1-4m_e^2/s}$. 
 Our  semi-analytical integration approach with physically
accessible observables as integration variables  may be  used 
to set benchmarks and to control the numerical precision to  more
than four digits.  
Their choice  is constrained by the observables we 
want to predict, notably the angular distribution
and certain cross-section asymmetries.
Basically we follow the approach proposed in
 { \cite{Bardin:1977qa,Passarino:1982zp}} 
and extend the required formulae to the massive fermion case. 

{There is not too much found in the literature for the
radiative production of massive fermion pairs.
Thus, we will discuss the kinematical details with some care, since they define
the integration boundaries of our numerical integration program.}

{
In (\ref{eqn:momenta}) and (\ref{eqn:momentau}) we defined $t$ and $u$
for two-particle production.
With the additional photon in the final state, we have to be more
specific and will use the following definitions:
\ba
\label{def-T}
T &=& 2p_4 q_3,
\\ \label{def-U}
U &=& 2p_1 q_3.
\ea
Additionally, the following invariants will be used:}
\ba\label{invariantsspri}
s'&=&(q_2+q_3)^2 \,, 
\\\label{invariantsz}
Z_{1,2}&=&2p  p_{1,4}\,,
\\
V_{1,2}&=&2p  q_{2,3}.
\label{invariants}
\ea
The squares of all three-momenta {in the centre-of-mass system} can be
 expressed in terms of  a set of $\lambda$ functions: 
\ba
\label{lamd1}
{4}s~|\vec q_2|^2 &=
~\lambda_1~ \equiv &\lambda[(p_1+p_4)^2,(q_3+p)^2,q_2^2]=(s-V_2)^2-4m_t^2s
\,,\\
 {4}  s~|\vec q_3|^2  &=
~\lambda_2~ \equiv &\lambda[(p_1+p_4)^2,(q_2+p)^2,q_3^2]=(s'+V_2)^2-4m_t^2s
\label{v12}
\,,\\
\label{lamds}
  {4}s~|\vec p_1|^2 ~=~ {4}s~|\vec p_4|^2 &=
~\lambda_s~ \equiv &\lambda[(p_1+p_4)^2,{p_1^2,p_4^2}]=s^2-4 m_e^2 s
\,,\\
 {4}  s~|\vec p|^2&=
~\lambda_p~ \equiv &\lambda[(p_1+p_4)^2,(q_2+q_3)^2,p^2]=(s-s')^2,
\label{lambda}
\ea
where we use $\lambda(x,y,z)=x^2+y^2+z^2-2(xy+xz+yz)$
{ and the relation $4p_A^2|\vec p_B|^2 = \lambda[ (p_A+p_B)^2,
  p_A^2, p_B^2 ]|_{\vec p_A =0}$ for 
$p_A = (p_1+p_4)$  $= (\sqrt{s},0,0,0)$.} 

The phase space of three particles in the final state is  
five-dimensional.
This means that only five of the ten scalar products 
{(those introduced already: $s, T, U, s', Z_i, V_i$, plus
  $W_{1,2}=2p_{1,4} q_2$)} 
built from the five momenta are independent. 
In fact, the following relations hold 
\ba
s &=& s'+V_1+V_2
\label{substv1} = s'+Z_1+Z_2 =  V_2+W_1+W_2
\\
  &=& V_1+U+T =Z_1+W_1 +  U .
\ea
We  use the first of relations (\ref{substv1}) in order to substitute $V_1$
everywhere in favour of $V_2$  as already done in (\ref{v12}).
Since 
we do not consider transversally polarized initial particles,
an integration of the cross section over the corresponding rotation angle is 
trivially giving a factor $2\pi$ and
we are left with four non-trivial phase space variables.

For the calculation of the forward--backward asymmetry, the angle $\theta$ 
between the three-momenta of $\bar t$ and {$e^+$}
is used{, in accordance with (\ref{def-T})}. 
Further, the energies $E_t$, $E_{\bar t}$, and $E_\gamma$ are `good' variables.
As mentioned before, they can be 
 expressed in terms of  the invariants {$s'$ and
$V_2$:} 
\ba
\label{energies1}
E_{\bar t}&=&\frac{s'+V_2}{2\sqrt{s}},
\\\label{energies2}
E_\gamma&=&\frac{s-s'}{2\sqrt{s}},
\\\label{energies3}
E_t&=&\sqrt{s}-E_\gamma-E_{\bar t}=\frac{s-V_2}{2\sqrt{s}}.
\ea 
{
The scattering angle in the centre-of-mass system may now be expressed
by invariants:
\ba
T &=& \frac{s'+V_2}{2} - \frac{\beta_0\sqrt{\lambda_2}}{2} ~\cos\theta. 
\ea}
As will be seen later, {the two invariants $s'$ and $V_2$} also
  describe the angles 
between any pair of  final-state particles.
Therefore, we choose {them} to parametrize the phase space. 
Finally, the 
fourth integration variable
will be the azimuthal angle of the photon $\phi_\gamma$.

The coordinate system is chosen such that  the $\bar t$ moves along the 
$z$ axis and the beam axis is in the $y$--$z$ plane.
The four-momenta of all particles can then be written as follows:
\ba
\label{momenta1}
p_1&=&\frac{\sqrt{s}}{2}(1,0,-\beta_0\sin\theta,-\beta_0\cos\theta),
\\
\label{momenta2}
p_4&=&\frac{\sqrt{s}}{2}(1,0,\beta_0\sin\theta,\beta_0\cos\theta),
\\
\label{momenta3}
p  &=&E_\gamma (1,\sin\theta_\gamma\cos\phi_\gamma,
                \sin\theta_\gamma\sin\phi_\gamma,\cos\theta_\gamma),
\\
\label{momenta4}
q_3&=&(E_{\bar t},0,0,|\vec q_3|),
\\
q_2&=&p_1+p_4-p-q_3.
\label{momenta}
\ea
The $\phi_\gamma$ and $\theta_\gamma$ are the azimuthal and
 polar angles of the photon. 
The expression for $\cos\theta_\gamma$ (and also that for $|\vec q_3|$)
can be obtained from 
$(\vec p+\vec q_2)^2=(\vec q_3)^2$,
\ba
\cos\theta_\gamma
=\frac{\lambda_1-\lambda_2-\lambda_p}{2\sqrt{\lambda_p\lambda_2}}
=\frac{s'(s-s')-V_2(s+s')}{(s-s')\sqrt{\lambda_2}},
\label{energies}
\ea
and again depends only on $s'$ and $V_2$.
The differential bremsstrahlung cross section (\ref{crossint}) 
takes the form
\ba
d\sigma = 
\frac{1}{(2\pi)^5}\frac{1}{2s\beta_0}|{\cal M}|^2\cdot
\frac{\pi}{16s}\mbox{d}\phi_\gamma\mbox{d}s'\mbox{d}V_2\mbox{d}\cos\theta\nll
\equiv \frac{1}{(2\pi)^5}\frac{1}{2s\beta_0}|{\cal M}|^2\cdot
\frac{\pi s}{16}\mbox{d}\phi_\gamma\mbox{d}r\mbox{d}x\mbox{d}\cos\theta.
\label{space}
\ea
In the last step, 
dimensionless variables are introduced:
\ba
\label{dimenles}
x&=&\frac{V_2}{s},
\\
r&=&\frac{s'}{s},
\\
r_m&=&\frac{4m_t^2}{s}.
\ea
The integration boundaries are either trivial ($\phi_\gamma$ and
$\cos\theta$) or can be found from the 
condition 
that the three three-vectors $\vec p,\ \vec q_2,\ \vec q_3$ form a triangle
with the geometrical constraint
{$\cos^2\theta_\gamma \leq 1$}. 
The four integration variables vary
within the following regions:
\ba
\label{neworder1}
0\le &\phi_\gamma &\le 2\pi,
\\
\label{neworder2}
\frac{x}{2x+r_m/2}\left(1+x-\sqrt{(1-x)^2-r_m}\right) \le &1-r&\le
\frac{x}{2x+r_m/2}\left(1+x+\sqrt{(1-x)^2-r_m}\right),
\\
\label{neworder3}
0\le &x&\le 1-\sqrt{r_m},
\\
-1\le &\cos\theta &\le +1.
\label{neworder}
\ea
If the order of integrations over $r$ and $x$ is interchanged,
their boundaries are
\ba
\frac{1-r}{2}\left(1-\sqrt{1-\frac{r_m}{r}}\right) \le &x&\le
\frac{1-r}{2}\left(1+\sqrt{1-\frac{r_m}{r}}\right),
\\
r_m\le &r&\le 1.
\label{oldorder}
\ea

\begin{figure}[H]
\begin{minipage}[t]{7.8cm}{
\nobody\vspace*{1cm}
\begin{center}
\hspace{-3cm}
\mbox{\epsfysize=7.0cm\epsffile[0 0 450 450]{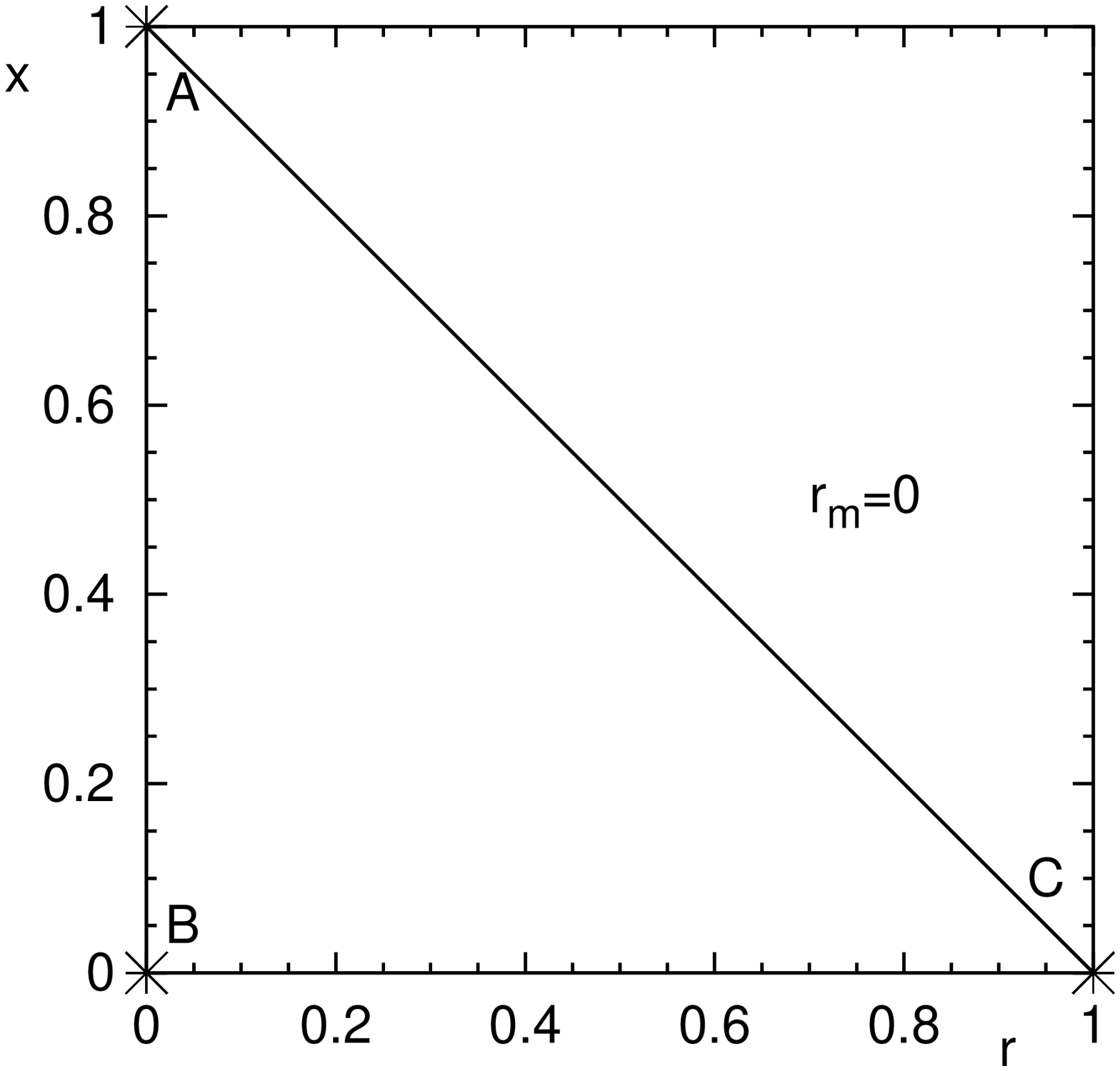}}
\vspace*{-1cm}
\end{center}
}\end{minipage}
\hspace{0.5cm}
\begin{minipage}[t]{7.8cm}{
\nobody\vspace*{1cm}
\begin{center}
\hspace{-3cm}
\mbox{\epsfysize=7.0cm\epsffile[0 0 450 450]{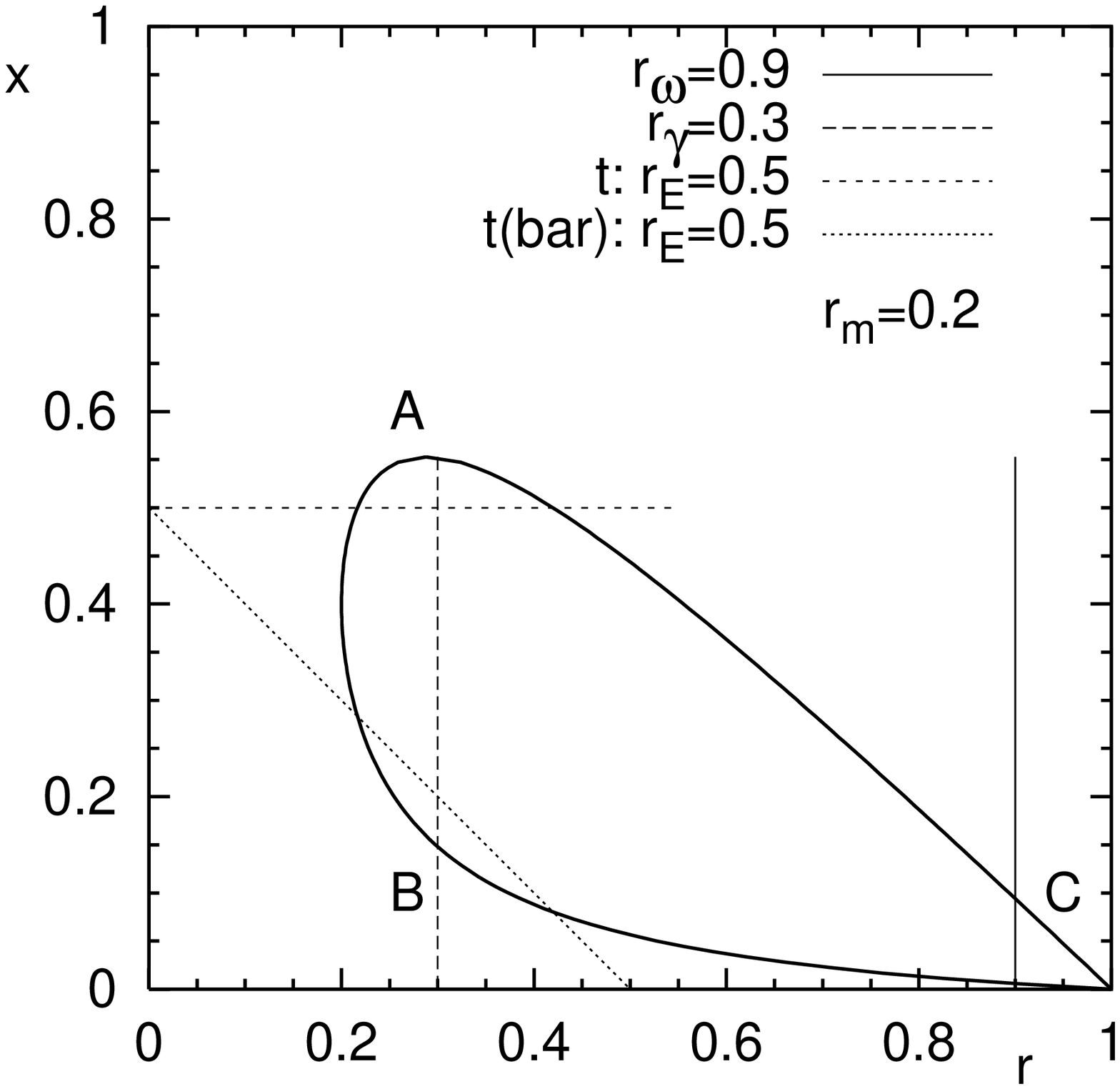}}
\vspace*{-1cm}
\end{center}
}\end{minipage}
\caption[]{\label{figurerx}
Kinematic region of $r$ and $x$ for (a) $r_m=4m_t^2/s=0$ and
(b)  $r_m=0.2$.   The 
energetic cuts are also shown; it is
$r_E=2E_{t}^{min}/\sqrt{s}$, 
$r_{\bar E}=2\bar E_{t}^{min}/\sqrt{s}$, 
$r_\omega=1-2\omega/\sqrt{s}=1-2E_{min}(\gamma)/\sqrt{s}$, 
$r_\gamma=1-2E_{max}(\gamma)\sqrt{s}$.}
\end{figure}
%

The kinematic regions of $r$ and $x$ are shown in Fig.~\ref{figurerx}(a) for massless ($r_m=0$) and in Fig.~\ref{figurerx}(b) for massive ($r_m\neq 0$) final
fermions. 
At the kinematic boundaries, the three-momenta 
$\vec p,\ \vec q_2,\ \vec q_3$ are parallel. 
Further, there are three special points where 
exactly one of the three three-momenta  vanishes
\ba
\label{acoord}
A&=&\left( \frac{\sqrt{r_m}}{2-\sqrt{r_m}},\ 1-\sqrt{r_m} \right),
\\ 
B&=&\left( \frac{\sqrt{r_m}}{2-\sqrt{r_m}},\ 
(1-\sqrt{r_m})~ \frac{\sqrt{r_m}}{2-\sqrt{r_m}} \right),
\label{abcoord}
\\
\label{abcoordc}
C&=&\left(1,0\right).
\ea
At point $C$ the soft photons are located.
{Section \ref{softint} is devoted to their treatment.}  
The $t$ ($\bar t$) are at rest in $A$ ($B$).
In the massless case, $r_m=0$, the three points $A,\ B$ and $C$ are located 
at the corners of the kinematic triangle,
$A=(0,1),\ B=(0,0),\ C=(1,0)$.
{From} (\ref{energies1})--(\ref{energies3}) it follows that 
the photon energy is maximal at the left edge, coinciding with the
$x$-axis;
 the fermion energy is maximal at the lower edge, coinciding with the $r$-axis;
and finally the energy of the anti fermion is maximal
at the third edge.

\subsubsection{Energy cuts}
Cuts on the energy of the final state particles are of importance for
two reasons: they are being applied in the experimental set-ups, and
for the photon we have to {identify} the soft photon terms in order to
combine them with virtual corrections for a finite net elastic cross section.  
The lower hard photon energy (being also the upper soft photon energy)
is
\ba
\label{egamdef}
\omega = E_\gamma^{\rm min}  .
\ea
All three energy cuts are deduced from 
(\ref{energies1}) to (\ref{energies3}).
The photon energy is related to $r$ by
\ba
\label{egamdef0}
r=1-2E_\gamma/\sqrt{s}
\ea
and the limits to be imposed are
\ba
\label{cut-e-phot}
r_\gamma = 1-2 E^{\rm max}_\gamma/\sqrt{s} \le r \le 
1-2\omega/\sqrt{s}=r_\omega.
\ea
Constraining the fermion energies leads to cuts on $x$:
\ba
\label{efermions}
r_{\bar E}-r \equiv 2E_{\bar t}^{\rm min}/\sqrt{s} -r \le x \le
1-r_E \equiv 1- 2E_{t}^{\rm min}/\sqrt{s}.
\ea
All the energy cuts are independent of the mass of the final fermions
and of $\cos\theta$.
They are illustrated in Fig.\ref{figurerx}.
{From} (\ref{efermions}) it can be seen that the derivatives of the 
kinematic border at points $A$ and $B$ in Fig.~\ref{figurerx}(b)
(for their definitions see (\ref{acoord}) and (\ref{abcoord})) are $0$
and $-1$.
\begin{figure}[t]
\begin{center}
\mbox{\epsfysize=7.0cm\epsffile[0 0 200 200]{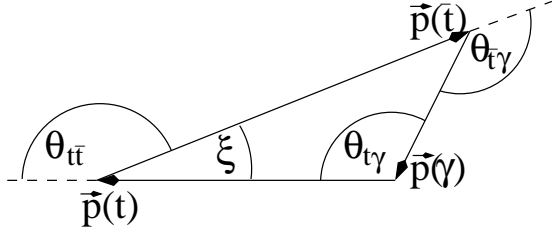}}
\vspace*{-1cm}
\end{center}
\caption[]{\label{figtriangle}
Triangle of the three-momenta of fermion, anti fermion and
photon.}
\end{figure}
\subsubsection{Angular cuts
\label{sec-angulcuts}}
 The scattering angle $\theta$

is the angle 
between $\bar t$ and $e^+$.
This angle is one of the integration variables and is constrained directly:
\bq
\label{coslim}
c_{\rm min} \le c \equiv \cos\theta \le c_{\rm max}.
\eq
Additional angular cuts deserve a study of the ($r,x$) parameter
space.
To be definite, we will always consider the kinematic bound of $r$ for
an arbitrarily chosen value of $x$.

The directions of the final-state particles define three angles
$\theta_{t\bar t}$,  $\theta_{t\gamma}$, and $\theta_{\bar t\gamma}$;
as shown In Fig.~\ref{figtriangle}.
The acollinearity angle $\xi$ is defined by
\bq
\xi=\pi-\theta_{t\bar t}.
\label{xidef}
\eq
The condition 
$\xi\ll 1$ restricts the events  to a Born-like kinematics: 
the fermions are back to back and 
only soft 
photons or photons collinear to one of the final fermions are allowed.
Using the above formulae, the acollinearity angle $\xi$ is 
 expressed in terms of  the invariants $x$ and $r$:
\bq
\cos\xi=\frac{\lambda_1+\lambda_2-\lambda_p}{2\sqrt{\lambda_1\lambda_2}}
= \frac{r(1+x)-x(1-x)-r_m}
{\sqrt{(1-x)^2(r+x)^2-r_m[(1-x)^2+(r+x)^2]+r_m^2}}.
\label{cosxi}
\eq
Equation (\ref{cosxi}) can readily be derived considering the scalar
product $q_2q_3$ or alternatively the triangle of the 
three-momenta of the final particles in {Fig.~\ref{figtriangle}}, {with account of  }the 
relations (\ref{lambda}) between $\lambda$ functions and absolute values 
of three-momenta.

\begin{figure}[t]
\begin{minipage}[t]{7.8cm}{
\nobody\vspace*{1cm}
\begin{center}
\hspace{-3cm}
\mbox{\epsfysize=7.0cm\epsffile[0 0 450 450]{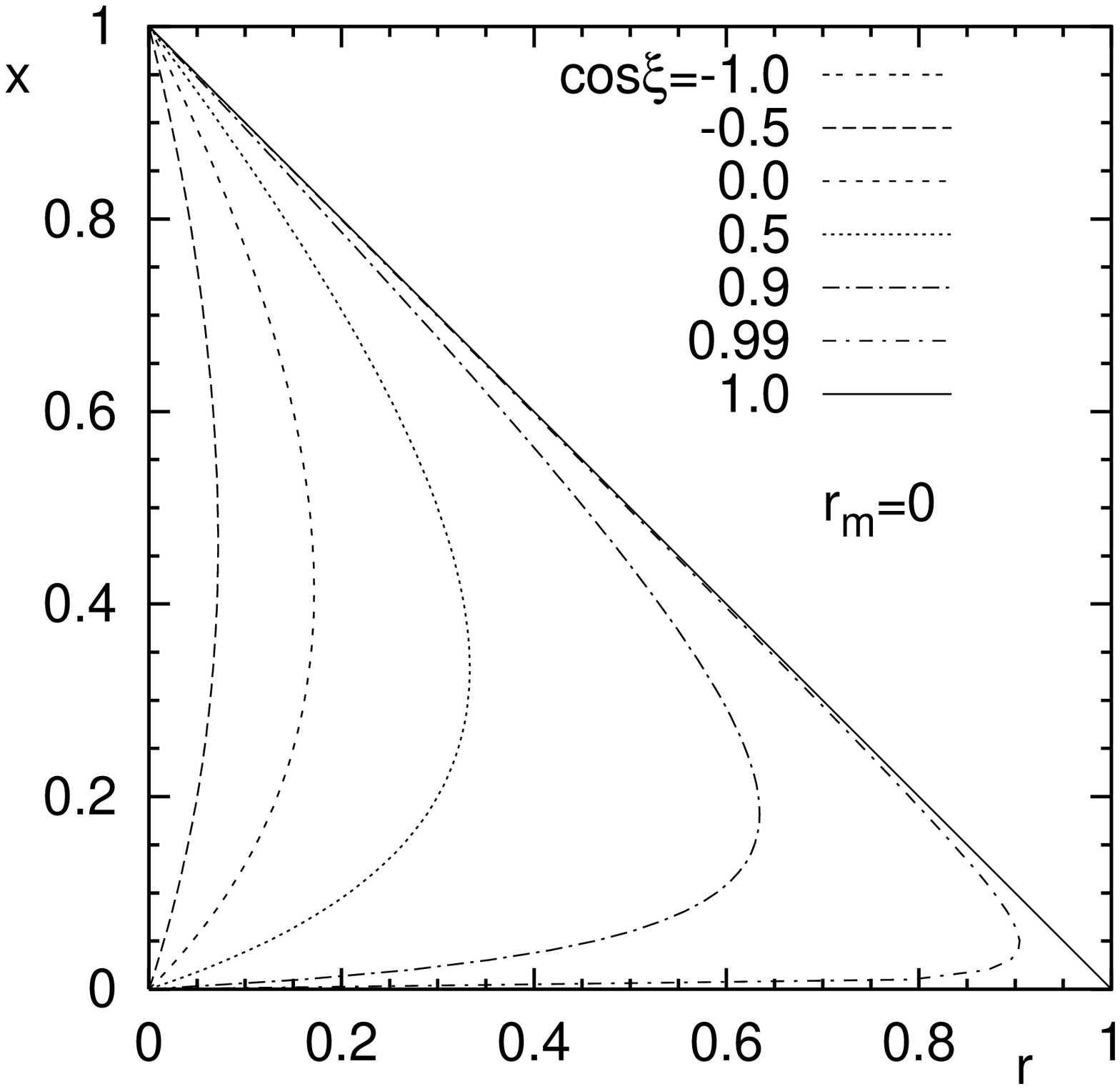}}
\vspace*{-1cm}
\end{center}
}\end{minipage}
\hspace{0.5cm}
\begin{minipage}[t]{7.8cm}{
\nobody\vspace*{1cm}
\begin{center}
\hspace{-3cm}
\mbox{\epsfysize=7.0cm\epsffile[0 0 450 450]{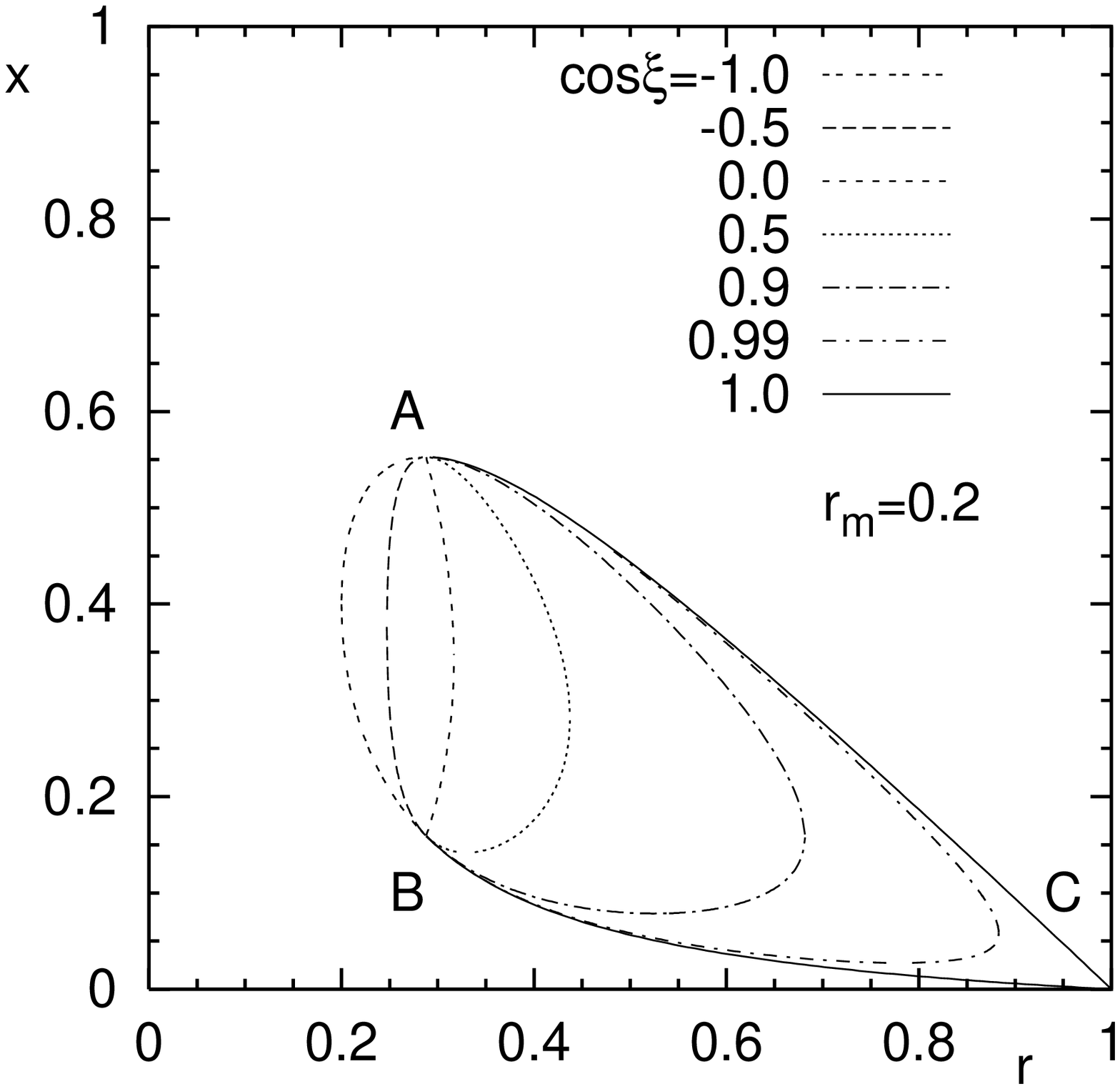}}
\vspace*{-1cm}
\end{center}
}\end{minipage}
\caption
{\label{fig-acol}
The kinematic region of $r$ and $x$ for different values of the acollinearity
angle $\xi$ for (a) $r_m = 4m_t^2/s=0$ and (b) $r_m$ = 0.2.
}
\end{figure}
%
{For} massless fermions, Eq.~(\ref{cosxi}) is much 
simpler
{and describes} a hyperbola with a symmetry axis rotated
by an angle of $-\pi/8$ relative to the $r$ axis.
The kinematic regions for different values of the acollinearity angle
are shown in Fig.~\ref{fig-acol}.
All lines intersect at the points $A$ and $B$.
For moderate cuts on the maximum acollinearity angle, the kinematic area is 
only constrained for values of $x$ above the point $B$, i.e. for 
$x>\sqrt{r_m}(1-\sqrt{r_m})/(2-\sqrt{r_m})$.
In this case only the lower bound of $r$ is changed.
The constraint to the kinematic region acts in a way similar to a cut to 
hard photons.
This is clear from the topology of events with high acollinearity:
the fermion and anti fermion fly approximately in one direction and must
recoil against a hard photon.
For stronger acollinearity cuts, constraints of the kinematic area arise 
also for values of $x$ below the point $B$.
In this case the allowed range for $r$ splits in two regions.
The first region extends from the lower kinematic border to the smaller 
solution of (\ref{cosxi}), while the second region extends from 
the larger solution of (\ref{cosxi}) to the upper 
kinematic border.

The analytic treatment of the acollinearity cut for the {\em massless}
case was also discussed in 
\cite{Montagna:1993mf,Christova:1999cc,Bardin:1999yd,Jack:2000as}.

In a similar way as explained for the acollinearity angle, the two angles 
$\theta_{t\gamma}$ and $\theta_{\bar t\gamma}$ can be  expressed in terms of  the 
two invariants $r$ and $x$:
\ba
\label{costeta5}
\cos\theta_{t\gamma}&=&\frac{\lambda_2-\lambda_1-\lambda_p}
{2\sqrt{\lambda_1\lambda_p}} 
= 
\frac{r(1+x)-(1-x)}
{(1-r)\sqrt{(1-x)^2-r_m}},
\\ 
\cos\theta_{\bar t\gamma} \equiv \cos\theta_{\gamma}
&=&\frac{\lambda_1-\lambda_2-\lambda_p}{2\sqrt{\lambda_2\lambda_p}}
=
{
\frac{-x(1+r)+r(1-r)}{(1-r)\sqrt{(x+r)^2-r_m}}
}.
\label{costeta6}
\ea 
Although
physically the situation for a cut on $\theta_{t\gamma}$ is equivalent to
a cut on $\theta_{\bar t\gamma}$,
the symmetry is broken because we had to make a choice between $V_1$
and $V_2$.  
The constraint on $\theta_{t\gamma}$ leads to {a quadratic
  equation} in $x$ {\it and} $r$,
while that for
$\theta_{\bar t\gamma}$ is quadratic in $x$ and of fourth order in $r$.
In the massless case, the constraints to $\theta_{t\gamma}$
 become a
{bi}linear equation in $r$ and $x$,
{describing} a hyperbola with a symmetry axis, which is rotated
by an angle $\pi/4$ relative to the $r$ axis.
The massless limit of (\ref{costeta6}) leads to 
a constraint on
$\theta_{\bar{t}\gamma}$, which is linear in $x$ and quadratic in $r$,
{describing} a hyperbola with a symmetry axis, which is rotated
by an angle of $\pi/8$ relative to the $x$ axis.

\begin{figure}[t]
\begin{minipage}[t]{7.8cm}{
\nobody\vspace*{1cm}
\begin{center}
\hspace{-3cm}
\mbox{\epsfysize=7.0cm\epsffile[0 0 450 450]{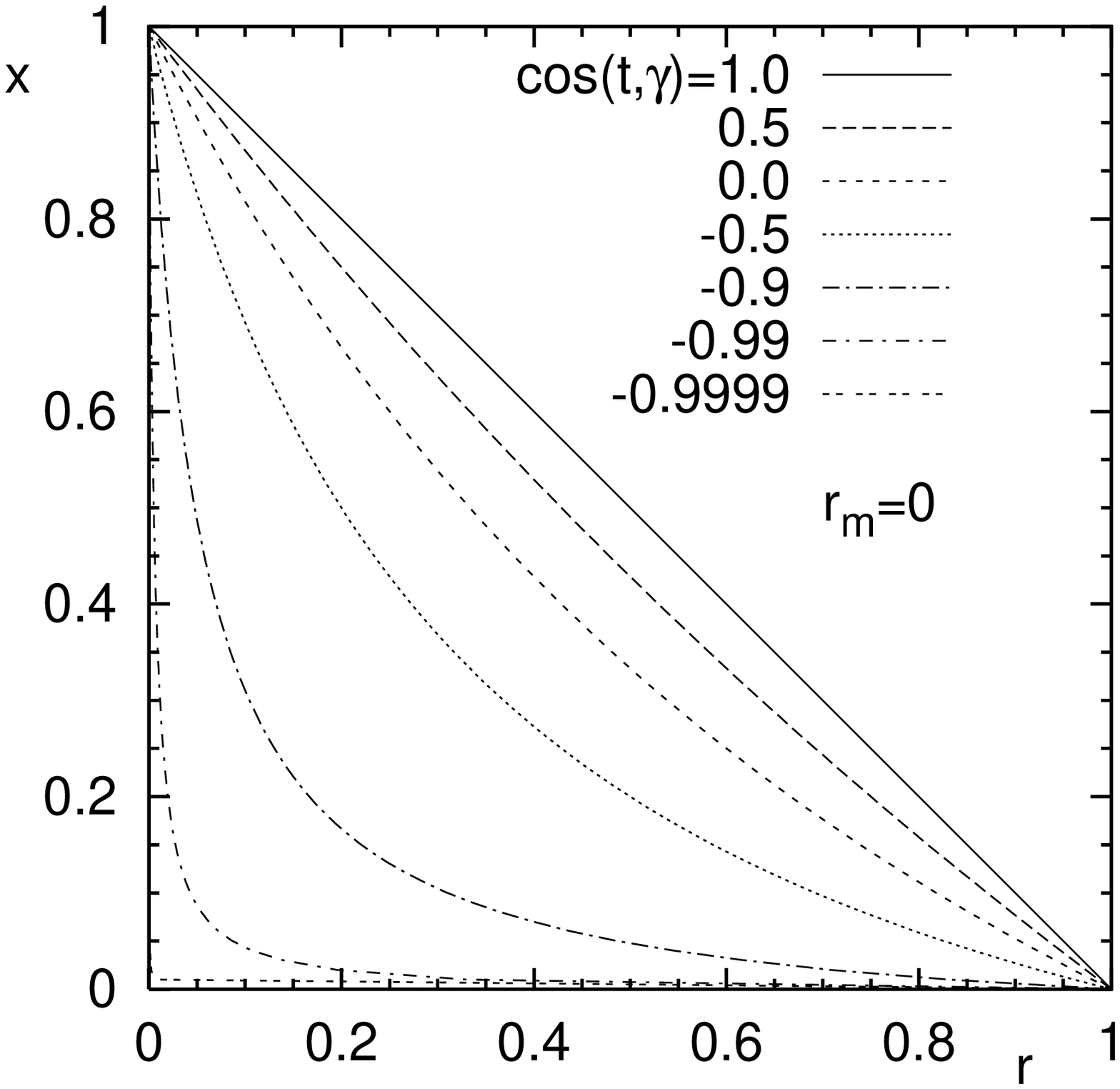}}
\vspace*{-1cm}
\end{center}
}\end{minipage}
\hspace{0.5cm}
\begin{minipage}[t]{7.8cm}{
\nobody\vspace*{1cm}
\begin{center}
\hspace{-3cm}
\mbox{\epsfysize=7.0cm\epsffile[0 0 450 450]{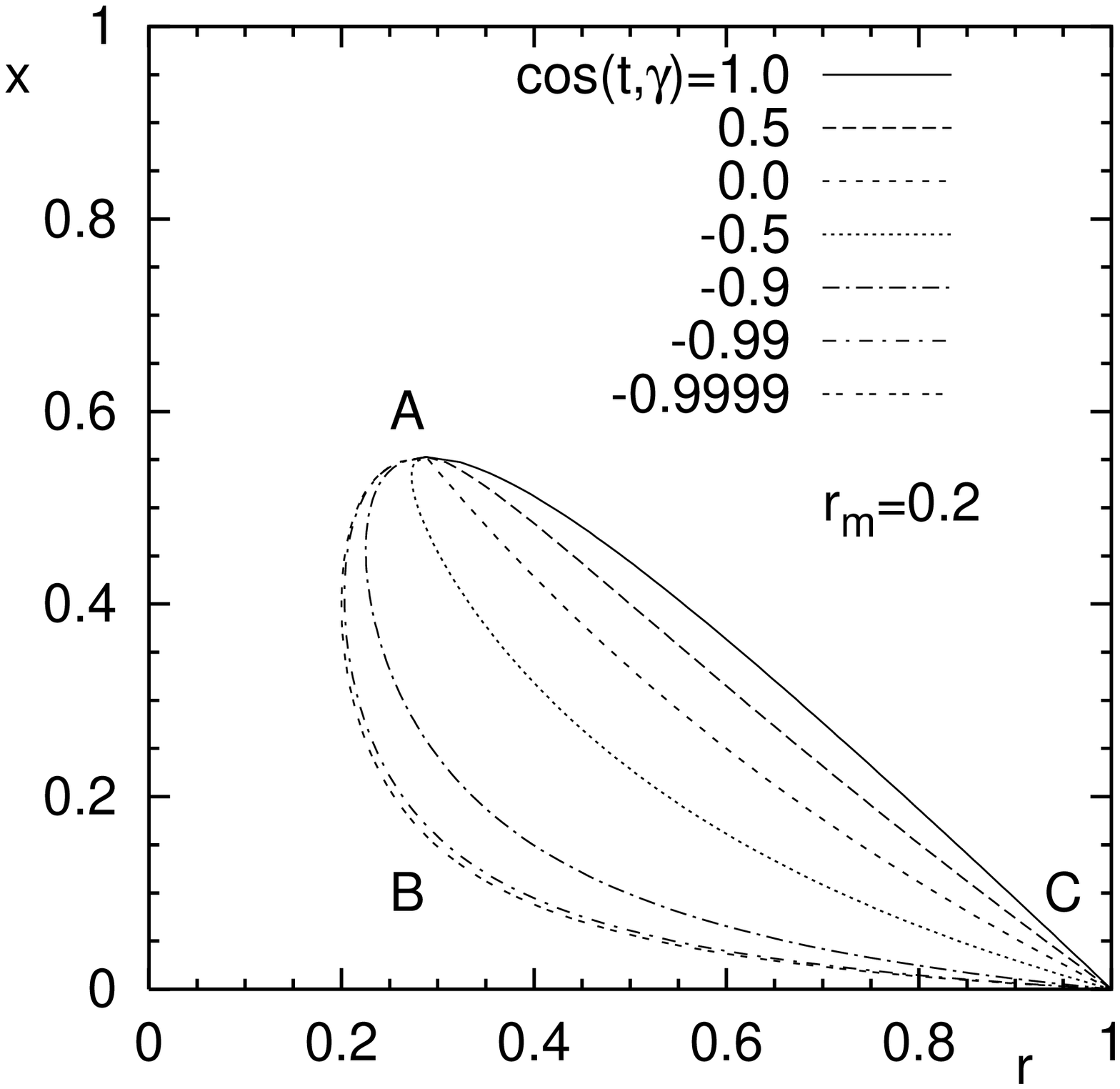}}
\vspace*{-1cm}
\end{center}
}\end{minipage}
\caption{\label{fig-tgam}
The kinematic region of $r$ and $x$ for different values of the angle 
$\theta_{t\gamma}$ and
for (a) $r_m = 4m_t^2/s=0$ and (b) $r_m = 0.2$.}
\end{figure}
%

The kinematic regions for different values of the angle $\theta_{t\gamma}$
are shown in Fig.~\ref{fig-tgam}.
All lines intersect at the points $A$ and $C$.
The exclusion of events with small angles $\theta_{t\gamma}$ excludes regions
near the edge $x=1-r$. 
These kinematic regions correspond to events with large anti fermion energies,
(compare Fig.~\ref{figurerx}).
Technically, a constraint of the angle $\theta_{t\gamma}$ from below 
changes the upper bound of $r$ in the kinematic region for a fixed $x$.
The lower bound of $r$ is unchanged by this cut.

\begin{figure}[t]
\begin{minipage}[t]{7.8cm}{
\nobody\vspace*{1cm}
\begin{center}
\hspace{-3cm}
\mbox{\epsfysize=7.0cm\epsffile[0 0 450 450]{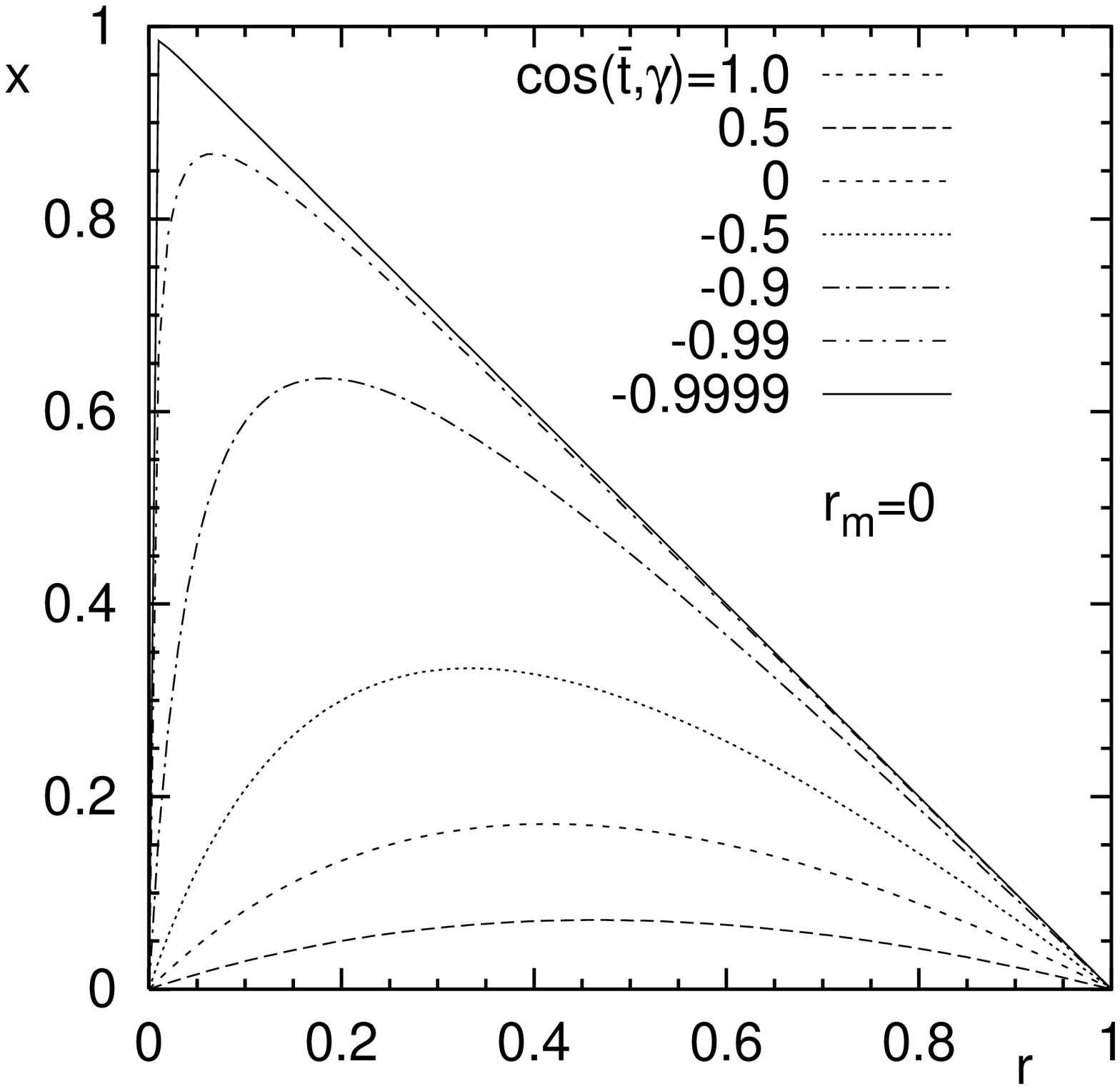}}
\vspace*{-1cm}
\end{center}
}\end{minipage}
\hspace{0.5cm}
\begin{minipage}[t]{7.8cm}{
\nobody\vspace*{1cm}
\begin{center}
\hspace{-3cm}
\mbox{\epsfysize=7.0cm\epsffile[0 0 450 450]{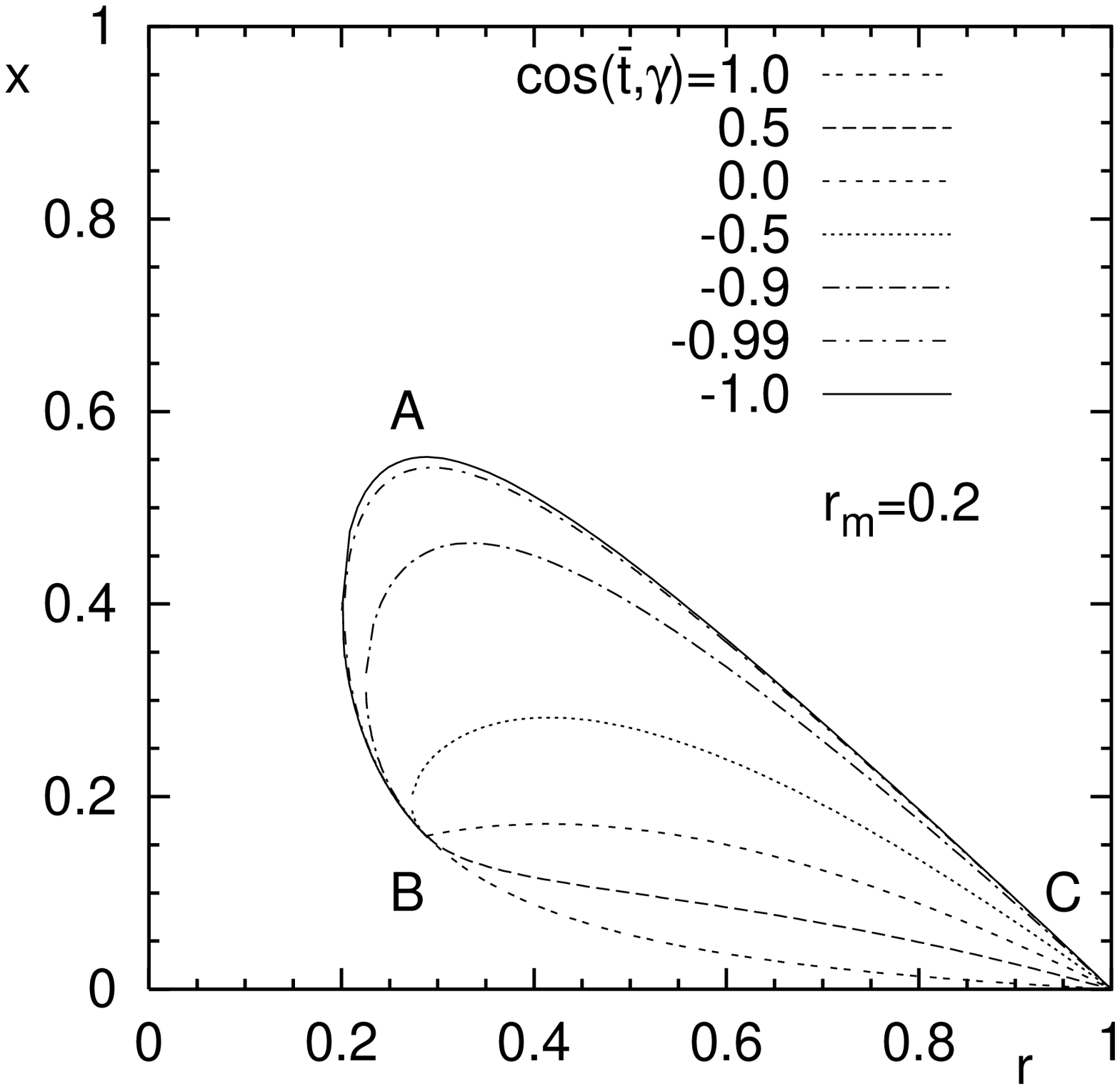}}
\vspace*{-1cm}
\end{center}
}
\end{minipage}
\caption{
\label{fig-antitgam}
The kinematic region of $r$ and $x$ for different values of the angle 
$\theta_{{\bar t}\gamma}$ and
for (a) $r_m = 4m_t^2/s=0$ and (b) $r_m = 0.2$.}
\end{figure}
%

Finally, the kinematic regions for different values of the angle 
$\theta_{\bar t\gamma}$ are shown in Fig.~\ref{fig-antitgam}.
All lines intersect at the points $B$ and $C$. 
The exclusion of events with small angles $\theta_{\bar t\gamma}$ excludes 
regions near the $r$ axis.
These kinematic regions correspond to events with large fermion energies,
(compare Fig.~\ref{figtriangle}).
Technically, a constraint of the angle $\theta_{\bar t\gamma}$ from below 
affects the kinematic region only for some $x$ below a certain value.
For these $x$ and for {\em massless} fermions,
the integration region of $r$ is split into two regions.
The first region extends from the lower kinematic border to the smaller 
solution of (\ref{costeta6}), and the second one extends from the 
larger solution of (\ref{costeta6}) to the upper kinematic 
border.
{For massive fermions, the
cutting out of }small angles $\theta_{\bar t\gamma}$ 
changes only the lower bound of $r$ as far as $x$ is below the point $B$.
For a harder cut on $\theta_{\bar t\gamma}$, the kinematic region is 
also affected for values of $x$,  larger than the $x$ coordinate of
the point $B$.
For a fixed $x$ and finite $m_t$, two regions of $r$ near the kinematic
border are then allowed, while some region in the ``middle'' is cut out by 
the constraint.
This is similar to the situation in the massless case.

\subsection{Radiative differential cross sections}
For massless fermions, {typically a threefold analytical integration}
of the radiative 
contributions to fermion pair production with realistic cuts {may be
performed, see 
 \cite{Christova:1999cc,Bardin:1999yd,Montagna:1993mf} and references
quoted therein}.
For massive pair production, everything becomes non-trivial and, in
the end, we
decided to perform only the first integration analytically, that over
$\phi_\gamma$.
This leaves three integrations at most for a numerical treatment.
Our practice proved that the accuracy and speed are absolutely
satisfactory  for our needs of calculating benchmarks and physics case
studies.

For this reasons, and in order to make everything well-defined, we now
have to collect the singly analytically integrated contributions to be
used in  a subsequent numerical calculation. 

We will use a notation for the couplings with some flexibility not
needed in the Born case:
{\allowdisplaybreaks
\ba
\label{vssp}
V(s,s')&=&\frac{1}{2}\Re e \Biggl\{
\sum_{m,n}[\chi_m(s)\chi_n^*(s')+\chi_m(s')\chi_n^*(s)]
\nobodyfrac
\nll &&
\times~
[v_e(m)v_e^*(n)+a_e(m)a_e^*(n)]\cdot [v_f(m)v_f^*(n)+a_f(m)a_f^*(n)]
\Biggr\} ,
\\
\label{assp}
A(s,s')&=&\frac{1}{2}\Re e\Biggl\{
\sum_{m,n}[\chi_m(s)\chi_n^*(s')+\chi_m(s')\chi_n^*(s)]
\nobodyfrac
\nll &&
\times~
[v_e(m)a_e^*(n)+a_e(m)v_e^*(n)]\cdot [v_f(m)a_f^*(n)+a_f(m)v_f^*(n)]\Biggr\},
\\
\label{cssp}
C(s,s')&=&2\Re e\Biggl\{
\sum_{m,n}[\chi_m(s)\chi_n^*(s')+\chi_m(s')\chi_n^*(s)]
\nobodyfrac
\nll &&
\times~
[v_e(m)v_e^*(n)+a_e(m)a_e^*(n)]\cdot a_f(m)a_f^*(n)\Biggr\},
\\
\label{csspp}
\hat C(s,s')&=&\Re e\Biggl\{
\sum_{m,n}[\chi_m(s)\chi_n^*(s')-\chi_m(s')\chi_n^*(s)]
\nobodyfrac
\nll &&
\nobodyfrac
\times~
[v_e(m)a_e^*(n)+a_e(m)v_e^*(n)]\cdot [v_f(m)a_f^*(n)-a_f(m)v_f^*(n)]\Biggr\},
\ea
}
where we use $v_f(\gamma)=Q_f$, $a_f(\gamma)=0$, $v_f(Z)=v_f$, and
$a_f(Z)=a_f$ and
\ba
\label{chizal}
\chi_Z(s)&=&\frac{s}{s-M_Z^2+i\Gamma_ZM_Z},
\\ 
\label{chiaal}
\chi_\gamma(s)&=&1.
\ea
With these conventions, the Born cross section becomes
\ba
\label{born-al}
\frac{d\sigma_{\rm Born}(s)}{dc}&=&\frac{\pi\alpha^2N_c\beta}{2s}
\left[V(s,s)(2-\beta^2+c^2\beta^2)+ 2c\beta A(s,s)
-\frac{1-\beta^2}{2}C(s,s)\right].
\ea

The cross section for $e^+e^-\rightarrow t\bar t\gamma$ subdivides in the 
gauge-invariant subsets of initial-state radiation, final-state radiation and
the interference between them.
Explicit expressions 
 for the totally differential cross section  
may be found in \cite{Jack:2000as}.
The integration over $\phi_\gamma$ is not too complicated and in fact
we could simply use existing tables of integrals
\cite{Bardin:1977qa,Passarino:1982zp,Jack:2000as}.    
This first integration is unaffected by the cuts discussed and 
has to be performed with 
an exact treatment of both $m_e$ and $m_t$.

The cross section for {\it initial-state radiation} after the integration
over $\phi_\gamma$ is
{\allowdisplaybreaks
\ba
\frac{d^3\sigma_{ini}}{dcds'dV_2}&=&
\frac{\alpha^3N_cQ_e^2}{2ss'^2}\left\{V(s',s')\left[
-2\frac{m_e^2\zeta_1}{\sqrt{D_1}^3}(2T^2-2Ts'+s'^2+2m_t^2s')
\right.\right.
\nll 
&&\left.\left.
-2\frac{m_e^2\zeta_2}{\sqrt{D_2}^3}(2U^2-2Us'+s'^2+2m_t^2s')
\right.\right.
\nll 
&&\left.\left.
+\frac{2s'}{s-s'}
\left(\frac{1}{\sqrt{D_1}}+\frac{1}{\sqrt{D_2}}\right)
(T^2-Ts'+U^2-Us'+s'^2+2m_t^2s')
\right.\right.
\nll 
&&\left.\left.
+\frac{1}{\sqrt{D_1}}(-2Us'+ss'+s'^2+2m_t^2(s+s'))
+\frac{1}{\sqrt{D_2}}(-2Ts'+ss'+s'^2+2m_t^2(s+s'))
\right.\right.
\nll 
&&\left.\left.
-2s'-4m_t^2\rule{0ex}{4ex}\right]
\right.
\nll 
&&\left.
+A(s',s')s'\left[
-2\frac{m_e^2\zeta_1}{\sqrt{D_1}^3}(s'-2T)
+2\frac{m_e^2\zeta_2}{\sqrt{D_2}^3}(s'-2U)
\right.\right.
\nll 
&&\left.\left.
+\frac{2s'}{s-s'}
\left(\frac{1}{\sqrt{D_1}}+\frac{1}{\sqrt{D_2}}\right)(U-T)
-\frac{1}{\sqrt{D_1}}(s+s'-2U)
+\frac{1}{\sqrt{D_2}}(s+s'-2T)
\right]
\right.
\nll 
&&\left.
+C(s',s')m_t^2\left[
2s'm_e^2\left(\frac{\zeta_1}{\sqrt{D_1}^3}+
\frac{\zeta_2}{\sqrt{D_2}^3}\right)
-\frac{2s'^2}{s-s'}
\left(\frac{1}{\sqrt{D_1}}+\frac{1}{\sqrt{D_2}}\right)
\right.\right.
\nll 
&&\left.\left.
-(s+s')\left(\frac{1}{\sqrt{D_1}}
+\frac{1}{\sqrt{D_2}}\right)
+2\right]
\right\},
\ea
with
\ba
\zeta_{1,2}&=&
\frac{s-s'}{2}\left(1
\pm\beta_0cc_t\right),
\\
c_t&=&\frac{V_1(s+s')-s(s-s')}{(s-s')\sqrt{\lambda_2}},
\\
\beta_0&=&\sqrt{1-\frac{4m_e^2}{s}},
\\
\label{eqn:c1c2}
D_{1,2}  = \frac{1}{\lambda_2}C_{1,2}
        &=&\frac{1}{\lambda_2}
         \left\{\frac{1}{4}\left[\nobodyfrac 2ss'-(V_2+s')(s+s')
\pm c\beta_0(s-s')\sqrt{\lambda_2}\right]^2
\right.\nll
&&\left. +~4m_e^2\left[\nobodyfrac
        s'V_2(s-s'-V_2)-(s-s')^2m_t^2\right]\right\}. 
\ea
}

The cross section for {\it final-state radiation} after the integration
over $\phi_\gamma$ is
{\allowdisplaybreaks
\ba
\frac{d^3\sigma_{fin}}{dcds'dV_2}&=&
\frac{\alpha^3N_cQ_f^2}{2s^3}\left\{V(s,s)\left[
-\frac{2m_t^2}{V_1^2}(-2UT+s^2+2m_t^2s)
\right.\right.
\nll &&\left.\left.
-\frac{2m_t^2}{V_2^2}(-2(UT+U\zeta_2+T\zeta_1+\zeta_{12})+s^2+2m_t^2s)
\right.\right.
\nll &&\left.\left.
\frac{4m_t^2}{V_1V_2}\left(\nobodyfrac U(T+\zeta_2)+T(U+\zeta_1)-2m_t^2s\right)
-\frac{2s}{V_1V_2}(2UT+t\zeta_2+T\zeta_1+\zeta_{12}-2s^2)
\right.\right.
\nll &&\left.\left.
+\frac{s}{V_1}(V_2-4m_t^2)+\frac{s}{V_2}(V_1-2s-4m_t^2)\right]
\right.
\nll &&\left.
+A(s,s)s\left[
\frac{2m_t^2}{V_1^2}(T-U)+\frac{2m_t^2}{V_2^2}(T+\zeta_2-U-\zeta_1)
\right.\right.
\nll &&\left.\left.
+\frac{1}{V_1V_2}(s'-2m_t^2)(2U+\zeta_1-2T-\zeta_2)
+\frac{U-T}{V_1}+\frac{U+\zeta_1-T-\zeta_2}{V_2}\right]
\right.
\nll &&\left.
+C(s,s)2m_t^2\left[
m_t^2s\left(\frac{1}{V_1^2}+\frac{1}{V_2^2}\right)
+\frac{1}{V_1V_2}(\zeta_{12}-ss'+2m_t^2s)\right]\right\} \,,
\ea
with
\bq
\zeta_{12}=\zeta_1\zeta_2-\frac{1}{8}(s-s')^2\beta_0^2(1-c^2)(1-c_t^2).
\eq
}

The cross section for the {\it interference between initial- and 
final-state radiation} after the integration over $\phi_\gamma$ is
{\allowdisplaybreaks
\ba
\frac{d^3\sigma_{int}}{dcds'dV_2}&=&
\frac{\alpha^3N_cQ_eQ_f}{2s^2s'}\left\{V(s,s')\left[
\frac{1}{\sqrt{D_1}V_1}(s-U)(2T^2-2Ts'+s'^2+2m_t^2s'-2UT+s^2
-2m_t^2s)
\right.\right.
\nll &&\left.\left.
-\frac{1}{\sqrt{D_2}V_1}(s-T)(2U^2-2Us'+s'^2+2m_t^2s'-2UT+s^2
-2m_t^2s)
\right.\right.
\nll &&\left.\left.
+\frac{1}{\sqrt{D_2}V_2}(s'-U)(2T^2-2Ts+s^2+2m_t^2s-2UT+s'^2
-2m_t^2s')
\right.\right.
\nll &&\left.\left.
-\frac{1}{\sqrt{D_1}V_2}(s'-T)(2U^2-2Us+s^2+2m_t^2s-2UT+s'^2
-2m_t^2s')
\right.\right.
\nll &&\left.\left.
+\frac{1}{V_1}\left( (\zeta_1-\zeta_2)s+(U-T)(3s-s'-4m_t^2)\right)
\right.\right.
\nll &&\left.\left.
-\frac{1}{V_2}\left( (\zeta_1-\zeta_2)(s'+4m_t^2)+(U-T)(3s'-s+4m_t^2)\right)
\right.\right.
\nll &&\left.\left.
+\left(\frac{1}{\sqrt{D_2}}
-\frac{1}{\sqrt{D_1}}\right)
\left(\nobodyfrac s^2+s'^2+2m_t^2(s+s')\right)\right]
\right.\nll &&\left.
+A(s,s')s\left[
 \frac{1}{\sqrt{D_1}V_1}(s-U)[-2Ts'+s'^2+s(U-T)]
\right.\right.
\nll &&\left.\left.
+\frac{1}{\sqrt{D_2}V_1}(s-T)[-2Us'+s'^2+s(T-U)]
+\frac{1}{\sqrt{D_2}V_2}(s'-U)[-2Ts+s^2+s'(U-T)]
\right.\right.
\nll &&\left.\left.
+\frac{1}{\sqrt{D_1}V_2}(s'-T)[-2Us+s^2+s'(T-U)]
\right.\right.
\nll &&\left.\left.
+\frac{1}{V_1}\left( 2ss'+2m_t^2(V_2-2s)\right)
+\frac{1}{V_2}\left(-2ss'+2m_t^2(V_1-2s)\right)
\right.\right.
\nll &&\left.\left.
+\frac{1}{\sqrt{D_1}}(s'T-sU)
+\frac{1}{\sqrt{D_2}}(s'U-sT) +2(s+s')+4m_t^2\right]
\right.\nll &&\left.
+C(s,s')m_t^2\left[
(s+s')\left(-\frac{1}{\sqrt{D_1}V_1}(s-U) 
+\frac{1}{\sqrt{D_2}V_1}(s-T)
-\frac{1}{\sqrt{D_2}V_2}T
+\frac{1}{\sqrt{D_1}V_2}U
\right)
\right.\right.
\nll &&\left.\left.
+(\zeta_1-\zeta_2)\left(\frac{1}{V_2}-\frac{1}{V_1}\right)\right]
\right.\nll &&\left.
+\hat C(s,s')m_t^2(s+s')\left[\frac{1}{V_1}+\frac{1}{V_2}\right]
\right\} \,.
\ea
}


\subsection{Soft photon corrections}
\label{softint}
The four-dimensional integration of the bremsstrahlung contributions
is divergent in the soft-photon part of the phase-space and is treated
 in $d$ dimensions.
One starts from a reparametrization of the photonic phase-space part
with Born-like kinematics for the matrix element squared.
To obtain a soft photon contribution we
have to take the terms of the bremsstrahlung amplitude 
without $p^0 \equiv E_{\gamma} \leq \omega$ in the numerators. 
In 
 this limit, $s'$ approaches  $s$ and  the soft contribution
to the differential cross section takes the  form
\ba
\label{dsoft}
\frac{d \sigma^{soft}}{d\cos\theta} =   
\frac{\alpha}{\pi} \delta^{soft} ~
\frac{d  \sigma^{Born}}{d\cos\theta}  
\ea
with
\ba
\label{delsoft1}
\delta^{soft}
 &=& 4 \pi^2 \int 
\frac{d^3 \vec p}{(2\pi)^3 2E_{\gamma}} \left[ 
Q_e \left(\frac{2 p_4}{Z_2} - \frac{2 p_1 }{Z_1} \right) + 
Q_t \left(\frac{2 q_2}{V_1} - \frac{2 q_3 }{V_2} \right)  \right]^2 
\theta(\omega -E_{\gamma})
\nll
&=&  
\frac{1}{4 \,\pi} \int 
\frac{d^3 \vec p}{E_{\gamma}^3} \theta(\omega - E_{\gamma})
I^{soft}
\ea
and
\ba
\frac{I^{soft}}{4E_{\gamma}^2}&=&
    Q_e^2 \left( \frac{ m_e^2}{Z_1^2} + \frac{ m_e^2}{Z_2^2} 
- \frac{s-2 m_e^2}{Z_1 Z_2} \right) 
+~ Q_e Q_t \left(  \frac{T}{Z_1 V_1}
+ \frac{T}{Z_2 V_2} - \frac{U}{Z_1 V_2}
- \frac{U}{Z_2 V_1} \right) 
\nonumber\\&&
+~ Q_t^2 \left( \frac{ m_t^2}{V_1^2} + \frac{ m_t^2}{V_2^2} 
- \frac{s-2 m_t^2}{V_1 V_2} \right) .
\ea
The scalar products have to be taken according to Born kinematics,
i.e. the expressions (\ref{invariantsz}) and (\ref{invariants}) become
\ba
\label{scalprod}
Z_1 &=& 2 p p_1  = 2 E_{\gamma} \left[p_1^0+|\vec{p}_1|\cos\theta_p\right],
\\
Z_2 &=& 2 p p_4  = 2 E_{\gamma} \left[p_4^0-|\vec{p}_4|\cos\theta_p\right],
\\
V_1 &=& 2 p q_2  = 2 E_{\gamma} \left[q_2^0+|\vec{q}_2|\cos\theta_q\right],
\\
V_2 &=& 2 p q_3  = 2 E_{\gamma} \left[q_3^0-|\vec{q}_3|\cos\theta_q\right].
\ea
From here we see that $I^{soft}$ is constructed to be independent of
$E_{\gamma}$.
Substitute now, { with $d=4-2\epsilon$, $\epsilon<0$,}
\ba
\label{delsoft2}
\delta^{soft} 
&\rightarrow&
\frac{(2\pi\mu)^{2\epsilon}}{4\pi} \int_0^{\omega}
E_{\gamma}^{{-(1+2\epsilon)}}d E_{\gamma} \int d\Omega_{(d-2)}I^{soft}
\nll
&=&
\frac{1}{2} \int_{-1}^1 d\xi \left[
{
{\cal P}_{IR}}
+ \ln\frac{\omega}{\mu} + \frac{1}{2} \ln(1-\xi^2)
 \right] I^{soft}.
\ea
We introduce the abbreviation for the infrared divergence
\ba
\label{pirdef3}
{\cal P}_{IR}
=
- \frac{1}{2\epsilon} + \frac{\gamma_E}{2} -\ln(2\sqrt{\pi}) .
\ea
The infrared divergence can also be regularized by introducing a
finite photon mass $\lambda$:
\ba
\label{pirdef}
P_{IR} -\ln\mu = \ln\frac{1}{\lambda}.
\ea

The last integral over $\xi = \cos\theta_{\gamma}$ is trivial for the
products $Z_iZ_j$ and $V_iV_j$, since they contain only
one angle;  one may thus identify either $\xi =
\cos\theta_p$ or $\xi = \cos\theta_q$. 
In the 
initial--final interference, one may introduce a Feynman parameter 
\ba
\frac{1}{Z_iV_j} = \frac{1}{4} ~ \frac{1}{p p_i~ p q_j} 
= 
\frac{1}{4} ~ \int_0^1d \alpha \frac{1}{(pk_{ij})^2}
\ea
with
\ba
k_{ij} = \alpha p_i + (1-\alpha)q_j .
\ea
Then, $(pk_{ij})^2 = E_{\gamma}^2 s
(1 - \beta_{ij} \cos\theta_{ij})^2 /4 $ and identify now 
$\xi = \cos\theta_{ij}$. Further,
\ba
(1-\beta_{12})^2 = (1-\beta_{43})^2 = (1-\beta_{T})^2 
&=& \frac{4}{s}\left[\alpha(1-\alpha)T
  +\alpha^2m_e^2+(1-\alpha)^2m_t^2\right] ,
\\
(1-\beta_{13})^2 = (1-\beta_{42})^2 = (1-\beta_{U})^2 
&=& \frac{4}{s}\left[\alpha(1-\alpha)U
  +\alpha^2m_e^2+(1-\alpha)^2m_t^2\right] . 
\ea
The final result is
\ba
\label{dsigsoft2}
\delta^{soft} &=& Q_e^2~ \delta^{soft}_{ini} 
+ Q_e Q_t~ \delta^{soft}_{int} + Q_t^2~ \delta^{soft}_{fin} 
\ea
with
\ba
\label{delsini}
\delta^{soft}_{ini}(m_e,\omega,\epsilon,\mu) &=&
2~\left(P_{IR}+\ln\frac{2\omega}{\mu}\right)
\left[
-1+\frac{s-2 m_e^2}{s\beta_0}\ln\left(\frac{1+\beta_0}{1-\beta_0}\right)
\right]
\nll &&
+~\frac{1}{\beta_0}\ln\left(\frac{1+\beta_0}{1-\beta_0}\right)
-
\frac{s-2 m_e^2}{s\beta_0}
\Biggl[
\frac{1}{2}\ln^2\left(\frac{2\beta_0}{1-\beta_0}\right)
\nll&&
+~\mbox{Li}_2(1)+\mbox{Li}_2\left(\frac{\beta_0-1}{2\beta_0}\right)
+\mbox{Li}_2\left(\frac{2\beta_0}{\beta_0+1}\right)
\Biggr] ,
\\
\label{delsofin}
\delta^{soft}_{fin}(m_t,\omega,\epsilon,\mu) &=&
\delta^{soft}_{ini}(m_t,\omega,\epsilon,\mu),
\\
\label{delsoint}
\delta^{soft}_{int}(m_e,m_t,\omega,\epsilon,\mu) &=&
2\left(P_{IR}+\ln\frac{2\omega}{\mu}\right)
\left( 
\frac{T}{\sqrt{\lambda_T}}
\ln\frac{T+\sqrt{\lambda_T}}{T-\sqrt{\lambda_T}} 
-\frac{U}{\sqrt{\lambda_U}}
\ln\frac{U+\sqrt{\lambda_U}}{U-\sqrt{\lambda_U}} 
\right)
\nonumber\\
&&
+~\frac{1}{2}\left[ T ~{\cal F}(T) - U ~ {\cal F}(U) \right],
\ea
and
\ba
\lambda_T &=& T^2-4 m_e^2 m_t^2,
\\
 {\cal F}(T) &=& - \frac{4}{s} \int_0^1 d\alpha \frac{1}{\beta_T(1-\beta_T^2)}
\ln\frac{1+\beta_T}{1-\beta_T},
\ea
and analogue definitions for $T \leftrightarrow U$.
We calculate the finite interference part given in (\ref{delsoint})
numerically, but have shown 
the agreement with Eq.~(3.64) of \cite{Beenakker:1989}:
\ba
\label{wim}
 T~ {\cal F}(T) - U ~ {\cal F}(U)
&=&
-2 \Biggl[
 \mbox{Li}_2\left(1-\frac{1-\beta}{1-\beta\cos\theta}\right)
+\mbox{Li}_2\left(1-\frac{1+\beta}{1-\beta\cos\theta}\right)
\nll &&
-~\mbox{Li}_2\left(1-\frac{1-\beta}{1+\beta\cos\theta}\right)
-\mbox{Li}_2\left(1-\frac{1+\beta}{1+\beta\cos\theta}\right)
\Biggr] .
\ea

\section{Results}

In this section we present the numerical results of the electroweak
one-loop calculation to the process \eett. We have performed a
fixed-order
 $\alpha$ calculation, i.e. no higher order corrections such as
photon exponentiation have been taken into account. \\

For the numerical evaluation we assume the following input values \cite{Fleischer:2002nn,Fleischer:2002rn,Fleischer:2002kg}:
\begin{align}
\begin{array}[r]{lll}
  \Gamma_Z = 2.49977 \GeV  \, \,,  & \alpha = \frac{e^2}{4\, \pi}
= 1 / 137.03599976 \,, & E_{\gamma}^{\max} = \sqrt{s}/10^5 \,, \\[2mm]
 \mw = 80.4514958 \GeV \, , & \mz = 91.1867 \GeV \,, & \mh = 120 \GeV \,,
 \\[1mm]
 m_e= 0.00051099907 \GeV \,, & m_t = 173.8 \GeV \,, & m_b = 4.7 \GeV \,,
 \\[1mm]
 m_{\mu} = 0.105658389 \GeV \,, & m_u = 0.062 \GeV \,, & m_d = 0.083
\GeV  \,, \\[1mm]
  m_{\tau} = 1.77705 \GeV \,, & m_c = 1.5 \GeV \,, & m_s = 0.215 \GeV \,.
\end{array}
 \end{align}
\ni
Two  packages, namely {\tt FF} \cite{vanOldenborgh:1991yc} and {\tt LoopTools}
\cite{Hahn:1998yk} have been  used for the numerical evaluation of the loop
integrals. \\

\begin{figure}[H]
 \centerline{\epsfysize=11.5cm\epsffile{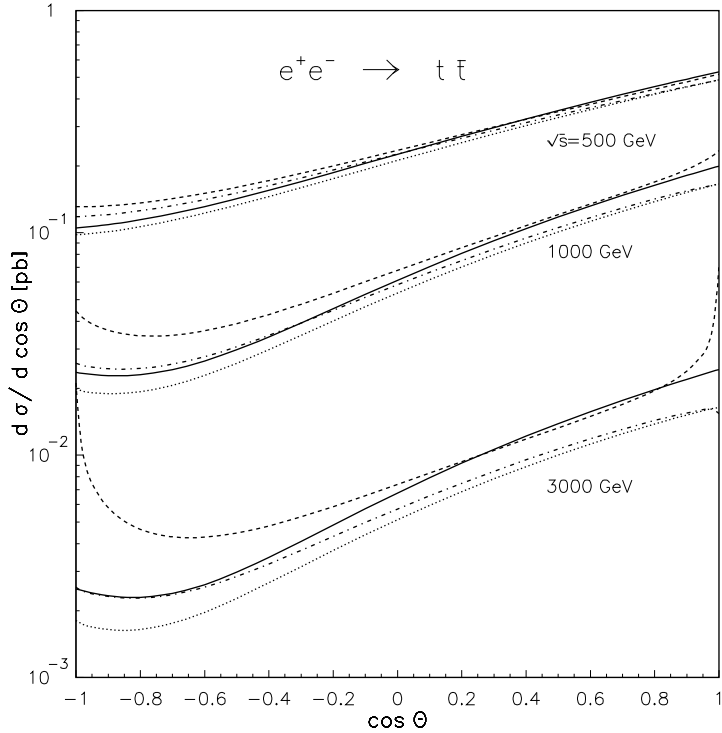}}
\vspace*{-0.7cm}
\caption{\label{eett}
Top-pair production:~~
Differential cross sections in Born approximation (solid lines), with
 full  electroweak corrections (dashed lines),  
with  an  $s'= 0.7 \, s$ cut (dash-dotted lines); 
 also shown:
pure weak corrections (dotted lines, photonic corrections and
running of $\alpha$ excluded); 
all for $\sqrt{s}$ = 0.5, 1, 3  TeV. }
\end{figure}

In Fig.~\ref{eett}, we 
present the differential cross section for various generic
values of $\sqrt{s}$. 

It can be seen that for rather high
centre-of-mass
 energies the characteristic features of a massive
fermion pair production become less prominent. At $\sqrt{s} = 3 $ TeV
the differential cross section of electroweak radiative corrections 
starts to exhibit collinear  mass  singularities at the
edges of phase 
space. 
Those are cured by applying 
a cut on $s^{\prime}$.  
In general it can be
seen that the effects of radiative corrections are more dramatic for
top-pairs produced close to the direction of the beam. For the TESLA
range of centre-of-mass energies,  backward scattered top quarks give
rise to slightly larger corrections than forward scattered ones \cite{Fleischer:2002rn }. For
higher energies this effect is more or less washed out.\\

In Table \ref{table1} to \ref{table3}  we present a complete set of form
factors entering the cross-section calculation. The form factors given
correspond to the minimal set of independent form factors possible
for a two-to-two process with two massless and two massive fermions in
the initial and final state respectively, and are defined with respect
to
the  `naturally'  arising form factors in Eq.~(\ref{eq:eqf61}). For
completeness we also give the corresponding Born form factors. The
numerical values given are obtained for a characteristic
centre-of-mass
 energy of $\sqrt{s} = 500 $ GeV and a fixed scattering angle
$\cos \theta = 0.7$. \\

\begin{table}[H]
\begin{center}
\begin{tabular}{|r|ll|ll|}

\hline  & &  & &  \\[-4mm]
 f.f. & \multicolumn{2}{c|}{Born}   & \multicolumn{2}{c|}{weak 1-loop contributions}\\
&\phantom{xxx} Re &\phantom{xxx} Im & \phantom{xxx} Re & \phantom{xxx}
Im
\\[1mm] 
\hline \hline 
 $\widehat{F}_1^{11}$ &$ -2.5092647\, 10^{-7}$&$
\phantom{-}6.0265891\, 10^{-12}$ &$\phantom{-}1.1805990\, 10^{-8}$&$-3.2896119\, 10^{-9}$ \\
 $ \,\widehat{F}_1^{15}$ &$\phantom{-}1.5620083\, 10^{-8}$&$-1.4732119\, 10^{-11}$ &$-1.0507915\, 10^{-8}$&$-8.4627303\, 10^{-9}$ \\
 $ \,\widehat{F}_1^{51}$ &$\phantom{-}5.6239963\, 10^{-8}$&$-5.3042857\, 10^{-11}$ &$-7.7050611\, 10^{-9}$&$-5.8986660\, 10^{-9} $\\
 $ \,\widehat{F}_1^{55}$ &$-1.3747972\, 10^{-7}$&$\phantom{-}1.2966433\, 10^{-10}
$ &$ -4.8821798\, 10^{-10}$&$\phantom{-}7.47501965\, 10^{-9}$\\
 $ \,m_t\,  \widehat{F}_3^{11}$ &$\phantom{-}0.0$&$\phantom{-}0.0$ &$\phantom{-}9.0882705\, 10^{-10}$&$-8.9067902\, 10^{-10}$\\
 $ \,m_t\, \widehat{F}_3^{51}$&$\phantom{-}0.0$&$\phantom{-}0.0$ &$-9.5315102\, 10^{-10}$&$\phantom{-}5.0995117\, 10^{-10}$\\
\hline
\end{tabular}
\end{center}
\vspace*{-0.5cm}
\caption[]{\label{table1}
Real and imaginary parts of the six independent form factors
$\widehat{F}_{i}^{jk} =\frac{e^2}{s} \, \bar{F}_i^{jk}$ for weak,
  non-photonic corrections to the process
\eett at $\sqrt{s} = 500$ GeV for a fixed scattering
angle $\cos \theta = 0.7 $. For reference we also give the
corresponding Born form factors. }
\end{table}

\begin{table}[H]
\begin{center}
\begin{tabular}{|r|ll|ll|}
\hline  & &  & &  \\[-4mm]
 f.f. & \multicolumn{2}{c|}{Born}   & \multicolumn{2}{c|}{weak 1-loop
contributions}\\
&\phantom{xxx} Re &\phantom{xxx} Im & \phantom{xxx} Re & \phantom{xxx}
Im
\\[1mm] 
\hline \hline 
 $\widehat{F}_1^{11}$
&$-6.2691435\, 10^{-8}$&$\phantom{-}3.5795119\, 10^{-13}$&$\phantom{-} 5.1890459\, 10^{-9}$&$-3.1198281\, 10^{-10}$\\
 $ \,\widehat{F}_1^{15}$ &$\phantom{-}3.8067962\, 10^{-9}$&$-8.7501891\, 10^{-13}$&$ -3.7904260\, 10^{-9}$&$-2.4180451\, 10^{-9}$\\
 $ \,\widehat{F}_1^{51}$ &$\phantom{-}1.3706334\, 10^{-8}$&$-3.1504975\, 10^{-12}$&$-3.0253509\, 10^{-9}$&$-1.8567076\, 10^{-9}$ \\
 $ \,\widehat{F}_1^{55}$ &$-3.3505411\, 10^{-8}$&$\phantom{-}7.7014546\, 10^{-12}$&$\phantom{-} 1.0657621\, 10^{-9}$&$\phantom{-}2.3331039\, 10^{-9}$\\
 $ \,m_t \,  \widehat{F}_3^{11} $&$\phantom{-}0.0$&$ \phantom{-}0.0$&$\phantom{-} 1.2278260\, 10^{-10}$&$-5.8408953\, 10^{-11}$\\
 $ \,m_t\, \widehat{F}_3^{51}$&$\phantom{-}0.0$&$ \phantom{-}0.0$&$  -9.9306672\, 10^{-11}$&$\phantom{-}4.2452469\, 10^{-11}$\\
\hline
\end{tabular}
\end{center}
\vspace*{-0.5cm}
\caption[]{\label{table2} Same as Table~\ref{table1}  for  $\sqrt{s}=1$ TeV. }
\end{table}


\begin{table}[H]
\begin{center}
\begin{tabular}{|r|ll|ll|}
\hline  & &  & &  \\[-4mm]
 f.f. & \multicolumn{2}{c|}{Born}   & \multicolumn{2}{c|}{weak 1-loop
contributions}\\
&\phantom{xxx} Re &\phantom{xxx} Im & \phantom{xxx} Re & \phantom{xxx}
Im
\\[1mm] 
\hline \hline 
 $ \, \widehat{F}_1^{11}$
&$-6.9644350\, 10^{-9}$&$\phantom{-}4.3540072\, 10^{-15}$&$\phantom{-}1.0150162\, 10^{-9}$&$\phantom{-}5.649373\, 10^{-12}$ \\
 $ \,\widehat{F}_1^{15}$ &$\phantom{-}4.1984821\, 10^{-10}$&$-1.0643459\, 10^{-14}$&$-6.7526508\, 10^{-10}$&$-3.4236400\, 10^{-10}$\\
 $ \,\widehat{F}_1^{51}$ &$\phantom{-}1.5116596\, 10^{-9}$&$-3.8321674\, 10^{-14}$&$-6.0751808\, 10^{-10}$&$-2.6754958\, 10^{-10}$\\
 $ \,\widehat{F}_1^{55}$ &$-3.6952823\, 10^{-9}$&$\phantom{-}9.3678104\, 10^{-14}$&$\phantom{-}3.5632400\, 10^{-10}$&$\phantom{-}3.4974067\, 10^{-10}$\\
 $ \,m_t\, \widehat{F}_3^{11}$ &$\phantom{-} 0.0$&$\phantom{-}0.0$&$\phantom{-}2.9895163\, 10^{-12}$&$-6.6708986\, 10^{-13}$\\
 $ \,m_t\, \widehat{F}_3^{51}$&$\phantom{-} 0.0$&$\phantom{-} 0.0$&$-2.4939160\, 10^{-12}$&$\phantom{-}9.1292861\, 10^{-13}$\\
\hline
\end{tabular}
\end{center}
\vspace*{-0.2cm}
\caption[]{\label{table3}Same as Table~\ref{table1}  for  $\sqrt{s}=3$ TeV. }
\end{table}

\begin{figure}[h]
\begin{center}
\nobody\hspace*{-1cm}
\mbox{\epsfysize=11.5cm\epsffile{
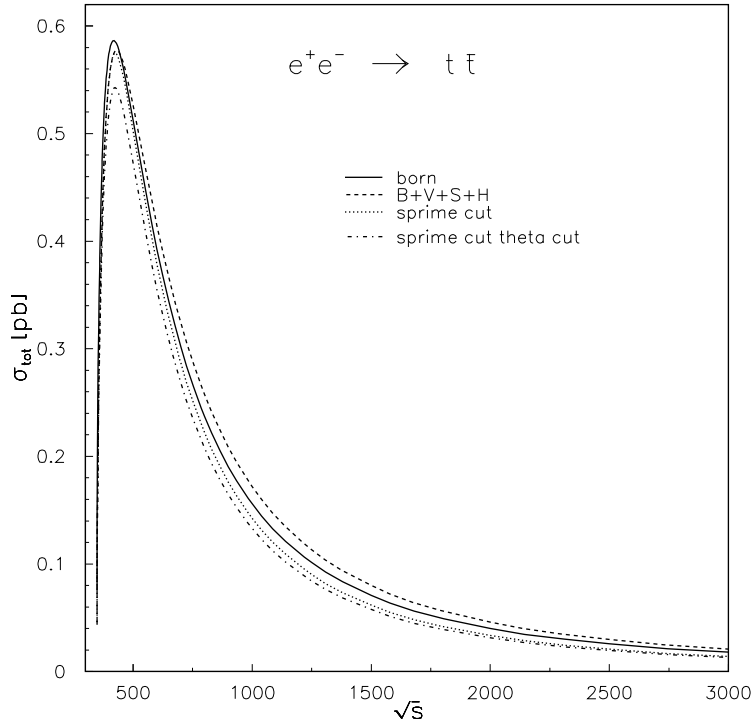}}
\vspace*{-1cm}
\end{center}
\vspace*{-0.7cm}
\caption[]{\label{figuretx}
Total cross-section for top-pair production as a function of $s$.
Born (solid lines), electroweak (dashed lines), electroweak with
$s'= 0.7 \, s$-cut (dotted lines)  and electroweak with  $s'=0.7 \, s$- and
$\cos\theta = 0.95$-cut (dash-dotted lines).}
\end{figure}

In Fig.~\ref{figuretx} we present the total integrated
cross section as a function of 
$\sqrt{s}$. 
From the previous discussion it is clear that the effect
of radiative corrections is less dramatic in the total cross section,
since the effects above and below the Born cross section are
averaged out.

%

Finally the forward--backward asymmetry of the total integrated cross
section can serve as a good observable to determine the effects of
radiative corrections. Towards higher energies, the effects become
distinctively. \\

\begin{figure}[H]
\begin{center}
\nobody\hspace*{-1cm}
\mbox{\epsfysize=11.5cm\epsffile{
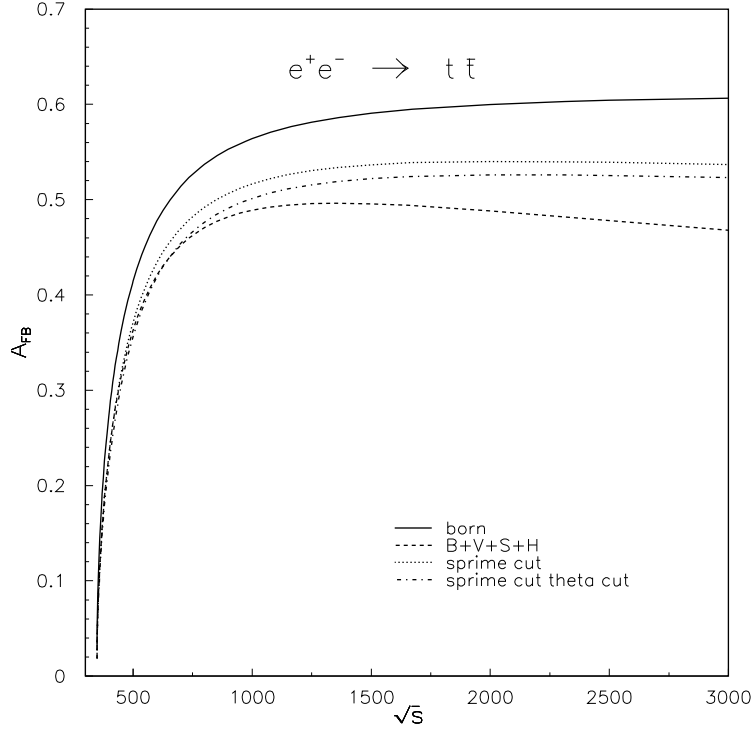}}
\vspace*{-1cm}
\end{center}
\vspace*{-0.7cm}
\caption[]{\label{figurefbx}
Forward--backward  asymmetry 
for top-pair production as a function of $s$.
Born (solid lines), electroweak (dashed lines), electroweak with
$s'$-cut (dotted lines)  and electroweak with  $s'= 0.7 \, s$- and
$\cos\theta = 0.95$-cut (dash-dotted lines).}
\end{figure}
%

In summary our calculation shows that for the next generation of
linear colliders with  centre-of-mass energies above $\sqrt{s}$ = 500 GeV,
electroweak radiative corrections modify the differential as
well as the integrated cross section within the experimental precision
of a few per mille. The package {\tt topfit} provides the means to
calculate those corrections and allows predictions for various
realistic cuts on the scattering angle as well as on the energy of the
photon.

\section*{Acknowledgements}

J.F. and A. L. would like to thank DESY Zeuthen for invitations
and all authors thank the Heisenberg-Landau projekt 
`New methods of computing massive Feynman integrals and
mass effects in the Standard Model' for financing visits to Dubna.
        
           \newpage
           \begin{appendix}
\section*{Appendices}
\label{chap:app}
\addcontentsline{toc}{section}{Appendices}
\def\thesubsection{\Alph{subsection}}
\def\theequation{\thesubsection.\arabic{equation}}
\setcounter{tocdepth}{0}
\setcounter{subsection}{0}
\setcounter{equation}{0}
\subsection{Translation of Tensor Decompositions
\label{app-extint}
}
On the left-hand side we give the  Passarino-Veltman 
functions,  used in \cite{vanOldenborgh:1991yc}, 
according to the
tensor decomposition 
of Feynman diagrams with respect to external momenta
as systematically introduced in 
\cite{Passarino:1979jh}.
On the right-hand side we follow the
corresponding notation in the {\tt LoopTools} package
\cite{Hahn:1998yk}, with $\lambda \neq 0$.
There are also sign differences reflecting different notions of metrics.
\begin{eqnarray}
   \label{eq:FFtoLT}
    C_0   & =& - C_0 \\
    C_{11} & =&  - C_1 - C_2 \\
    C_{12}  & =& - C_2 \\
 C_{24}  & =& C_{00} \\
 C_{21} & =& -C_{11} - 2\, C_{12} - C_{22} \\
  C_{22} & =& - C_{22} \\
C_{23}  & =& - C_{12} -C_{22} \\[2mm]
D_{11}  & =& D_1 + D_2 + D_3 \\
D_{12}  & =& D_2 + D_3 \\
D_{13}  & =& D_3 
\\
D_{21}  & =& D_{11} + 2\, D_{12} + 2\, D_{13} + 2\, D_{23} + D_{22} +
             D_{33} 
\\
 D_{22}  & =& 2 \, D_{23} + D_{22} + D_{33} 
\\
D_{23}  & =& D_{33} 
\\
 D_{24}  & =& D_{12} + D_{13} + 2\, D_{23} + D_{22} + D_{33}
 \\ 
D_{25}   & =& D_{13} + D_{23} + D_{33} \\
D_{26} & = & D_{23} + D_{33} \\
D_{27}  & =& - D_{00}
\end{eqnarray}


\subsection{Renormalization
\label{app-re}}
\setcounter{equation}{0}
A detailed formulation  of renormalization of fermion
pair production can be found in various textbooks, e.g. \cite{Bohm:2001yx}.
To complete the documentation of our calculation we
present  some relations resulting from
the application of an on-mass-shell renormalization{, closely following
\cite{Fleischer:1981ub}.} 
They had been used to derive the formulae given in Section \ref{sec-virtual}.

After the renormalization of the boson self-energies: we have to
use the following expressions:
\begin{eqnarray}
  \label{eq:effselfren}
  \Sigma_{ Z}^{\rm ren}(p^2) &=& \Sigma_{Z }(p^2) - {\rm Re}\,
\Sigma_{Z }(\mz^2)  \equiv   \Sigma_{Z }(p^2) - \delta \mz^2 \\
  \Sigma_{\gamma }^{\rm ren}(p^2) &=& \Sigma_{\gamma  }(p^2)  \\ 
  \Sigma_{\gamma Z}^{\rm ren}(p^2) &=& \Sigma_{\gamma Z }(p^2) 
\end{eqnarray}

The divergent parts    of those renormalized self-energies
 were   given in (\ref{ren-17}). 
 For the mixing angle renormalization ${\rm Re}
\, \Sigma_{ Z}(\mz^2) = \delta \mz^2 $ and ${\rm Re} \,
\Sigma_{ W}(\mw^2) = \delta \mw^2$ are needed:
 
\begin{eqnarray}
\label{ren-sw}
\delta \swto =  {\cwto} \,  \left(
\frac{\delta \mz^2}{\mz^2} - \frac{\delta \mw^2}{\mw^2} \right) \,.
\end{eqnarray}
Among the free parameters of the theory we have, only one
coupling constant $e$, using 
\begin{eqnarray}
  \label{eq:ZReImS}
g \sw = g' \cw = e = \sqrt{4\pi\alpha_{em}(0)}
\end{eqnarray}
The electric charge renormalization differs in pure QED and
electroweak theory
\begin{eqnarray}
\label{app-eqed}
e^{2,ren} &=& 4\pi\alpha_{em}(0) \left(1+2 \frac{\delta e}{e} \right)
\\
  \label{eq:deltaeQED2}
   \frac{\delta e}{e}^{\rm QED}  &=& -  \frac{1}{2} \, \delta Z_{\gamma}
   =  \frac{1}{2} 
\, \dd{p}  \Sigma_{\gamma}(p^2)_{\onshellA}
\\
\label{eq:deltaeweak}
    \frac{\delta e}{e} ^{\rm weak}  &=& \frac{1}{2} \,  \, \dd{p}
\Sigma_{\gamma}(p^2)_{\onshellA} - \frac{\sw}{\cw} \,
\frac{\Sigma_{Z\, \gamma}(0)}{\mz^2}
\end{eqnarray}
 The wav-function renormalization factor $Z_f$ is obtained from
the fermion self-energy $ \Sigma_f$, with
\begin{eqnarray}
  \label{eq:wavew}
 \Sigma_f(p)&=& A(p^2)+B(p^2)\,(\ps-m_f)+C(p^2)\,\ps\gamma_5\,
\end{eqnarray}
The resulting $Z$ factor is:
\begin{eqnarray}
 \label{eq:de3}
Z_f &=& 1+z_{a, f}+z_{b, f}~\gamma_5   \,.
\nll
&=& 1+B(m_f^2)+2m_fA^{\prime}(p^2)|_{m_f^2} + C(m_f^2) ~\gamma_5 \,.
\end{eqnarray}
For QED, the axial terms vanish, of course.
Explicitly,  the UV-divergent parts are given by :
\\
 
\begin{eqnarray}
\label{eq:de4}
 z_{a,f}^{UV} &  =& 
     -   \,\frac{e^2}{\swto} \frac{1}{\epsilon} \, \left(
   \frac{3}{8}\,     \frac{m_f^2}{\mw^2}    + \frac{1}{8} \frac{m_{f'}^2}{\mw^2} \right)
       - \,e^2\,  \frac{1}{\epsilon}   \, ( Q_f^2 + a_f^2 + v_f^2 ) \\
 z_{b,f}^{UV} & =& - \,\frac{e^2}{\swto} \frac{1}{\epsilon} \left(  -
\frac{1}{4} + \frac{1}{8} \frac{m_f^2}{\mw^2} - \frac{1}{8}
\,\frac{m_{f'}^2}{\mw^2}  \right)
       + \,e^2\, \frac{1}{\epsilon}   \, (   2\,a_f\,v_f ) \\
\label{eq:de8}
\frac{\delta e^{{\rm weak} \, ,UV}}{e} &=&  e^2  \frac{11}{6} ~\frac{1}{\varepsilon} 
 \\
\label{eq:de9}
\delta \swto^{UV} &=&  e^2  \left(\frac{41}{6} - \frac{21}{2}  \cwto
  +\frac{11}{3}  \cwfor \right)
                    \frac{1}{\swto} ~\frac{1}{\varepsilon}    = e^2 \left(\frac{41}{6} -
\frac{11}{3} \, \cwto  \right) \frac{1}{\epsilon} 
\end{eqnarray}
 with $f'$ denoting the isospin partner of $f$.

The above relations define the complete renormalization procedure
needed for our reaction.
A vertex renormalization, e.g. resulting from terms 
such as $e\bar{\Psi}\gamma_{\mu}A_{\mu}\Psi$ in the Lagrangian, traces back to $\delta e$ and
$Z_f$.   
{Explicit formulae may be found in the Fortran code \cite{FRW:2002sw}.}


\def \litwo {{\rm Li_2}}
\subsection{Infrared Divergences \label{chap:IR}}
\def\onshellm{\vphantom{\frac{1}{1}}_{\left|_{\scr{\,p^2=m^2}}\right.}}
\def \dd {{\rm d}}
\def \ni {\noindent}
\def \no {\nonumber}
\def \dd#1{\frac{\partial}{\partial \, {#1}^2}}
\setcounter{equation}{0}
{ The conventions of the one-loop functions and related
  ones are those used in the package {\tt LoopTools}
  \cite{Hahn:1998yk}.
In particular the normalization of the one-loop integration
is used as in the following simplest example:
\begin{eqnarray}
\label{eq:a0}
A_0(m^2) ~=~ \frac{(2\pi\mu)^{4-d}}{i\pi^2} \int \frac{d^d
  k}{k^2-m^2} 
&=&{-(4\pi\mu^2)^{2-\frac{d}{2}}
   \frac{1}{(m^2)^{1-\frac{d}{2}}} \Gamma(1-\frac{d}{2})} 
\nll
&=& m^2~\left[1-\ln\frac{m^2}{\mu^2} + \frac{1}{\epsilon}
  -\gamma_E+\ln(4\pi)\right] + {\cal O}(\epsilon).
\end{eqnarray}}

In the Fortran program, we leave the treatment of IR divergences to
the packages used for the calculation of one-loop integrals.
Additionally, we checked  analytically their cancellation. 
For this purpose, we isolated  them in the few IR-divergent scalar 
integrals contributing to the process \eett. 

One-loop infrared divergences are due to the exchange of a photon between
two massive particles, which occur also as external (on-shell) ones.

 Wave function renormalization yields  IR-divergent contributions
$DB_0$ and $DB_1$, the on-mass-shell derivatives of $B_0$ and 
$B_1$ (w.r.t. the external 
momentum squared). From 
\footnote{In {\tt FF}, there is no $DB_1$ 
  foreseen, while in {\tt LoopTools} this function was numerically unstable for
  $\lambda\neq0$. 
This might 
  be  improved now. }
With the representation
\begin{eqnarray}
\label{eq:b1red}
 B_1(p^2;m_1^2,m_2^2)=\frac{1}{2p^2}
            \left[(m_2^2-m_1^2-p^2)~B_0(p^2,m_1^2,m_2^2) + A_0(m_1^2) 
- A_0(m_2^2)\right]
\end{eqnarray}
one arrives at
\begin{eqnarray}
\label{eq:b2red}
DB_1(m^2;m^2,0) &\equiv& \dd{p} B_1(p^2;m^2,0)\onshellm \no \\
&=& \frac{1}{2 \, m^4}
\left[ - A_0(m^2) +m^2 \, B_0(m^2;m^2,0) - 2\, m^4 DB_0(p^2;m^2,0)
\onshellm  \right]  .
\end{eqnarray}
 The UV divergences cancel at the right-hand side and the IR divergence
is traced back to $DB_0$.
We mention for completeness that the similar function 
$DB_1(m^2;0,m^2) \equiv \dd{p} B_1(p^2\!=\!m^2;0,m^2)
= [  A_0(m^2) -m^2  B_0(m^2;0,m^2)] /(2m^4)$, arising from the
charged current self-energy with a massless neutrino, 
is finite.

An explicit calculation gives
\begin{eqnarray}
\label{eq:b3red}
DB_0(p^2;m^2,0)\onshellm 
& =& \frac{1}{\left(m^2\right)^{3-d/2}} ~~\frac{\Gamma(3-d/2)}{(d-3)(d-4)}   
{ (2{\sqrt{\pi}}\mu)^{4-d}}
\no \\
& =& { -}~ \frac{C_0(m^2,0,m^2;0,m^2,m^2)}{d-3},
\\ \label{c0fordb0}
{ C_0(m^2,0,m^2;0,m^2,m^2)} &\simeq& - \frac{1}{m^2} \ln \frac{ m}{\lambda} .
\end{eqnarray}

Assigning the loop-momentum $k$ to the photon line in the initial-  and
 final-state vertex diagrams ensures that the divergent part is
exclusively contained in one scalar three-point function $C_0$ : 
\begin{eqnarray}
\label{eq:C1red}
  C_0(m^2,s,m^2;0,m^2,m^2)  &=& 
\frac{{-1}}{s {\beta}}
\Biggl\{
\ln(y) \left[2\ln(1+y)-\frac{1}{2}\ln(y)-\ln\frac{\lambda^2}{m^2}   \right]
+\frac{\pi^2}{6} +2\litwo(-y)
\Biggr\} 
\no \\
\end{eqnarray}
with
\begin{eqnarray}
\label{eq:C4red}
y\equiv y(s,m,m) 
= \frac{\sqrt{1-4m^2/{ s}}-1}{\sqrt{1-4m^2/{ s}}+1}
= \frac{{\beta} -1}{{\beta} + 1} + i \varepsilon.
\end{eqnarray}
{{The finite, small photon mass $\lambda$ is defined according
    to (\ref{pirdef}).}} 

\begin{figure}[tbh]
\begin{center}
\mbox{\epsfysize=5.0cm\epsffile{
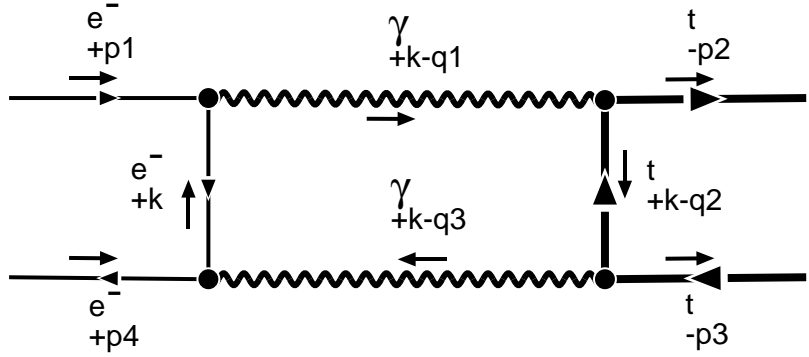}}
\vspace*{0cm}
\caption{\label{dianatt1012} {An infrared-divergent box diagram.}}
\end{center}
\end{figure}

\ni 
%
%
Finally, IR-divergent functions from the photonic box
diagrams, $D_0$, , $D_{\mu}$, and $D_{\mu \nu}$, have to be considered.
We indicate for the direct two-photon box, shown in Fig.~\ref{dianatt1012},
how the singularities can be isolated.

\noindent
The key ingredient for this method \cite{Passarino:1979jh} is the following identity:
\begin{eqnarray}
  \label{eq:IRboxsplit1}
D_0 &\propto &  \int  \, \frac{d^4 k}{\c{1} \, \c{2} \, \c{3} \, \ck{4}
} \\ \no
&& \h{-6}  
=\frac{-1}{s-m_1^2-m_3^2} 
\Biggl[
\int  
\frac{ 2 \, ( k -q_1) \, (k-q_3)  ~d^4 k}
{\c{1} \, \c{2} \, \c{3} \, \ck{4}} 
\\ \no
&&  - \frac{d^4 k}{\c{2} \, \c{3} \, \ck{4}} -
  \frac{d^4 k}{\c{1} \, \c{2} \ck{4} } \Biggr] .
\end{eqnarray}
 For \m{m_1 = m_3 = 0}, evidently 
the numerator of the first of the three terms makes it 
an IR-finite contribution and the other two are $C_0$ functions.
 To demonstrate more explicitely the procedure we select the above 
diagram (see Fig.~\ref{dianatt1012}) and obtain (see also
\cite{Fleischer:1989kj}): 
{\allowdisplaybreaks
\begin{eqnarray}
 \label{eq:IRboxsplitkk}
s~D_0 
&\propto & 
 s~ \int  \, \frac{d^4 k \,  }
{\co{1} \, 
\ensuremath{ [ (k - q_{2})^2 - m_{t}^2 ]} 
\, \co{3} \, 
\ensuremath{ [ k^2 - m_{e}^2 ]}} 
\nll
& \eqir& 
\int\frac{d^4 k}{
\ensuremath{ [ (k - q_{2})^2 - m_{t}^2 ]} 
\, \co{3} \,
\ensuremath{ [ k^2 - m_{e}^2 ]}
} 
+  
\int\frac{d^4 k}{\co{1} \, 
\ensuremath{ [ (k - q_{2})^2 - m_{t}^2 ]} 
\ensuremath{ [ k^2 - m_{e}^2 ]}
} 
 \no 
 \\
& \eqir& 
C_0(q_2^2,(q_3-q_2)^2,q_3^2;m_e^2,m_t^2,0) +
C_0(q_1^2,(q_1-q_2)^2,q_2^2;m_e^2,0,m_t^2)  \no 
\\
&=& 
2~C_0(t,m_t^2,m_e^2;m_e^2,m_t^2,0) .
\end{eqnarray}
}
Only one scalar function has to be calculated, and in the limit
of vanishing $m_e$ we find
\begin{eqnarray}
\label{eq:C3red}
{C_0(m_e^2,t,m_t^2;0,m_e^2,m_t^2)} &=&  
\frac{1}{T}\left[
-\ln\frac{m_e m_t}{T}\ln\frac{\lambda^2}{m_e m_t} + 
{\litwo}\left( 1- \frac{m_t^2}{T} \right)
-~\frac{1}{2}
\ln\frac{m_e^2}{T}\ln\frac{m_t^2}{T}
\right] .
\end{eqnarray}
\\
From the crossed box diagram, we get another function, $D_0$, with $t$
in Eq.~(\ref{eq:C3red}) replaced by $u$.
When combining virtual and soft corrections, the singularities of
these functions are cancelled
against the divergent parts of (\ref{delsoint}).

The vector and tensor functions may be treated quite similarly:
{\allowdisplaybreaks
\begin{eqnarray}
 \label{eq:IRboxlitkk}
s~D_{\mu} &\propto& \h{2}  \int  \, \frac{ k_{\mu}~ d^4k  }{\co{1} \, \c{2} \, \co{3} \, \ck{4}
}  
\nll
&  \eqir &
 \int\frac{ [(k_{\mu}-q_{3\, \mu})+q_{3\, \mu}]d^4k }{\c{2} \, \co{3} \,
\ck{4}} +  \int\frac{[k_{\mu} -q_{1\, \mu})+q_{1\, \mu}]d^4k}
{\co{1} \,\c{2} \ck{4} } \no 
\\
&  \eqir &\left(q_{1\, \mu} \, + q_{3\, \mu} \,
\right) C_0(m_e^2,t,m_t^2;0,m_e^2,m_t^2)  ,
\\ [5mm]
s~D_{\mu \nu} &\propto& \h{2}  \int  \, \frac{ k_{\mu} \,
  k_{\nu} d^4k}{\co{1} \, \c{2} \, \co{3} \, \ck{4}}  
\nll &  \eqir &
\left(q_{1\, \mu} \, q_{1 \, \nu} \,
  + q_{3\, \mu} \, q_{3\,
 \nu}\, \right)  C_0(m_e^2,t,m_t^2;0,m_e^2,m_t^2) .
\end{eqnarray}
}

To cross-check the result, we 
isolated the IR-divergent parts also with another approach,
where the
tensor integrals are reduced to scalar ones by means of
recurrence relations  \cite{Tarasov:1996br,Fleischer:1999hq}. 
The divergent contributions hidden in the tensor integrals manifest
themselves in the form of the three IR-divergent scalar
functions $C_0$ introduced above, {namely (\ref{c0fordb0}) 
for self-energies,  (\ref{eq:C1red}) for vertices,  and
(\ref{eq:C3red}) for boxes, correspondingly. 
}


           \end{appendix}
           \newpage

           \addcontentsline{toc}{section}{References}
\bibliographystyle{utphys_spires}
\bibliography{toppair}
\end{document}